\documentclass{emulateapj}

\usepackage{graphicx}
\usepackage{epsfig}
\usepackage{amsmath}

\newcommand{\htwo}{\,\hbox{$\rm{H_ 2}$}}

\newcommand{\hbeta}{\,\hbox{{H${\rm \beta}$}}}
\newcommand{\paa}{\,\hbox{{Pa${\rm \alpha}$}}}
\newcommand{\brg}{\,\hbox{{Br${\rm \gamma}$}}}
\newcommand{\brd}{\,\hbox{{Br${\rm \delta}$}}}
\newcommand{\hei}{\,\hbox{\ion{He}{1}}}

\newcommand{\oiii}{\,\hbox{[\ion{O}{3}]}}
\newcommand{\feii}{\,\hbox{[\ion{Fe}{2}]}}
\newcommand{\caviii}{\,\hbox{[\ion{Ca}{8}]}}
\newcommand{\sivi}{\,\hbox{[\ion{Si}{6}]}}
\newcommand{\HI}{\,\hbox{\ion{H}{1}}}
\newcommand{\sixi}{\,\hbox{[\ion{Si}{11}]}}

\newcommand{\tex}{\,\hbox{$T_{\rm ex}$}}
\newcommand{\msun}{\,\hbox{$M_{\odot}$}}

\newcommand{\kms}{\,\hbox{\hbox{km}\,\hbox{s}$^{-1}$}}
\newcommand{\degree}{\ensuremath{^\circ}}

\newcommand{\um}{\,\hbox{$\mu$m}}

\newcommand{\vcirc}{\,\hbox{$V_{c}$}}
\newcommand{\hst}{{\it Hubble Space Telescope}}

\shorttitle{A radio jet drives a molecular and atomic gas outflow within 1\,kpc$^2$ of the nucleus of IC5063}
\shortauthors{K. M. Dasyra et al.}

\submitted{  Received 2015 March 17; accepted 2015 September 22}
\journalinfo{The Astrophysical Journal}

\begin{document}

\title{A radio jet drives a molecular and atomic gas outflow in multiple regions within one square kiloparsec of the nucleus of the nearby galaxy IC5063}

\author{K. M. Dasyra$^{1,2}$, A. C. Bostrom$^{3}$, F. Combes$^{2}$, N. Vlahakis$^{1}$}

\affil{$^1$Department of Astrophysics, Astronomy \& Mechanics, Faculty of Physics, National and Kapodistrian University of Athens, Panepistimiopolis Zografos 15784, Greece \\
$^2$Observatoire de Paris, LERMA, 61 Av.\,de l'Observatoire, F-75014, Paris, France\\
$^3$University of Maryland, College Park, MD 20742, USA\smallskip}

\begin{abstract}

We analyzed near-infrared data of the nearby galaxy IC5063 taken with the Very Large Telescope SINFONI instrument. IC5063 is an elliptical galaxy that has a radio jet nearly aligned with the major axis of a gas disk in its center.  The data reveal multiple signatures of molecular and atomic gas that has been kinematically distorted by the passage of the jet plasma or cocoon within an area of $\sim$1 kpc$^2$. Concrete evidence that the interaction of the jet with the gas causes the gas to accelerate comes from the detection of outflows in four different regions along the jet trail: near the two radio lobes, between the radio emission tip and the optical narrow-line-region cone, and at a region with diffuse 17.8\,GHz emission midway between the nucleus and the north radio lobe. The outflow in the latter region is biconical, centered 240\,pc away from the nucleus, and oriented perpendicularly to the jet trail.  The diffuse emission that is observed as a result of the gas entrainment or scattering unfolds around the trail and away from the nucleus with increasing velocity. It overall extends for $\gtrsim$700\,pc parallel and perpendicular to the trail. Near the outflow starting points, the gas has a velocity excess of 600\kms\ to 1200\kms\ with respect to ordered motions, as seen in \feii , \paa , or \htwo\ lines. High \htwo\ (1-0) S(3)/S(1) flux ratios indicate non-thermal excitation of gas in the diffuse outflow.

\end{abstract}


\section{Introduction}
\label{sec:intro}

Cosmological simulations that required the implementation of black-hole feedback to regulate galaxy growth \citep[e.g.,][]{croton06,bower06} attributed a crucial role to radio jets. Jets can affect the interstellar medium (ISM) of galaxies by expelling, heating, and potentially compressing the gas. Heating was mainly invoked in the cosmological simulations because it prevents the accretion of new baryons from the cosmic web into galaxies, and because it disperses already accreted baryons into a hot, tenuous medium, in which cooling is inefficient \citep{maio07}. Heating has been observationally confirmed for the tenuous gas around Perseus A, from ripples in X-ray-based temperature and pressure maps \citep{fabian06}. It has also been proposed for the molecular gas in the outflow of 4C12.50, where the outflowing-to-ambient \htwo\ mass is $\gtrsim$30 times higher at 400\,K than at 25\,K \citep{dasyra12,dasyra14}. Thermally driven bubbles drive gas expansion and, thus, produce the same observable signatures as shocks and ram pressure: outflows. 

Several outflow detection experiments have been successfully carried out for radio galaxies. Absorption from accelerated \HI\ gas at the tip of radio hot spots was reported by \citet{morganti05}. Accelerated ionized gas close to the radio jet lobes was detected in optical wavelengths \citep[e.g.,][]{rupke11,shih13}. X-ray data revealed the presence of nuclear ultra-fast outflows, moving at about 10\% of the lightspeed, in radio-loud galaxies \citep{tombesi14}. Recently, massive outflows of molecular gas were detected in galaxies of different morphological types showing (weak or strong) radio emission  \citep[e.g.,][]{feruglio10,dasyra11,nesvadba11,alatalo11,alatalo14,garciaburillo14,mcnamara14,sakamoto14}. These detections were considered of particular importance because the molecular gas can probe star formation and because it can significantly contribute to the mass load of the outflow thanks to its high density. In some of these systems, the outflow entrained a few percent of the entire available reservoir. 

From a theoretical viewpoint, the detection of such massive outflows can be explained through the fragmentation and reformation of dense molecular clouds because of the passage of tenuous ionized gas around them. Once the tenuous gas flow is initiated, thermodynamic (Rayleigh-Taylor and Kelvin-Helmholtz)  instabilities act upon the surface of the molecular clouds, dispersing them, and facilitating their acceleration by ram pressure \citep[e.g.,][]{klein94,hopkins10,zubovas14}. Molecules reform down the mass-loaded flow thanks to cooling, which happens on very short timescales with respect to the flow travel time \citep{maio07}. 

Following these results and their interpretation, models treating radio jets as a heating source of the tenuous gas \citep[e.g,][]{sijacki06} were followed by models treating radio jets as a source of shock-induced energy dissipation upon the dense gas \citep[e.g.,][]{wagner11,gaibler12}. A major conclusion from the work of \citet{wagner11} is that the jet cocoon influences the disk not only along the jet propagation axis, but also 500\,pc (or more) in the transverse direction. This prediction needs to be tested in galaxies in which we can spatially resolve jets carving their way through disks. Such examples are rare according to the standard unification scheme of active galactic nuclei \citep[AGN; ][]{urry95}. One galaxy with this peculiarity is NGC 4258 \citep{cecil00,wilson01,krause07}. \citet{wilson01} associated diffuse X-ray emission 170\,pc above the disk with hot gas in the jet cocoon.

Another valuable example is that of IC5063, a nearby elliptical galaxy with a central gas disk and a radio jet that propagates for several hundred parsec before it scrapes out of the disk plane \citep{morganti98}. In radio data,  bright synchrotron emission is seen at its two lobes, near the bases of the narrow-line-region (NLR) ionization cones \citep{morganti98}.  At the two radio lobes,  as well as at a clump that is near the north-west edge of the NLR ionization cone, \citet{kulkarni98} detected strong line emission in \hst\ NICMOS narrow-band images that they attributed to gas excited by jet-related shocks. Strong polarization of the near-infrared (NIR) continuum light was seen at the north radio lobe by \citet{lopez13}. A fast neutral gas outflow was detected through \HI\ absorption in front of this lobe \citep{morganti98,oosterloo00}. A molecular outflow, which is associated or thought to be associated with the same region, has been reported by \citet{morganti13} for the cold gas seen in millimeter CO data, and by \citet{tadhunter14} for warm gas seen in NIR \htwo\ data. These authors proposed that the outflow is driven by the impact of jet plasma upon dense ISM clouds. 

We analyzed SINFONI data aiming to study the effects of the jet propagation on the atomic and, mainly, on the molecular ISM of IC5063. We adopted H$_0$=70 \kms\  Mpc$^{-1}$, $\Omega_{M}$=0.3, and $\Omega_{\Lambda}$=0.7 throughout our work.
 
\section{Data reduction}
\label{sec:data*}

The NIR data used in this work were taken from the archive of the European Southern Observatory (ESO). We reduced all available spectrophotometric data of IC5063 that were taken with the SINFONI integral-field unit of the Very Large Telescope (VLT). On-source observations of two hours, divided into 12 exposures of 600\,s, were carried out in the H band on June 6 and June 7 2005 (as part of the program 075.B-0348 A). Another two-hour-long observing run took place on May 22 and June 6 2012, using the H+K instrument setup. This time, 24 on-source exposures of 300\,s were taken (as part of the program 089.B-0971 A). All datasets were taken with a field of view (FOV) of 8.0$\arcsec $$\times$8.0\arcsec. The observations, which were carried out under seeing conditions of 1.0$\arcsec$$\pm$0.3$\arcsec$, did not benefit from (laser-guide-star-assisted) adaptive-optic corrections. 
 
 We started the data reduction by removing data rows affected by erroneous detector-hardcoded processing, using the IDL routine provided by ESO for this purpose. We then used the ESO Recipe Execution (EsoRex) version 3.10 tool  to generate a nonlinear-pixel map, a master dark frame, a master flatfield frame, and a bad-pixel map per night, using the routines sinfo\_rec\_detlin, sinfo\_rec\_mdark, and sinfo\_rec\_mflat, respectively. We then computed the optical distortion of the spectra of the multiple slitlets and the wavelength calibration solution with the aid of lamp exposures (using the routines sinfo\_rec\_distortion and sinfo\_rec\_wavecal). We ran the last pipeline routine, sinfo\_rec\_jitter, to reduce individual science frames of stars used as telluric/flux calibrators and of IC5063. This routine applied the previous calibration results to the science frames, and removed sky frames from the science frames. For the 2012 program data, one sky frame was assigned to two science frames because of the employed object-sky-object nodding strategy. 
 
 Once the science cubes were reconstructed, an atmospheric absorption correction was applied to them. To perform this correction, we used the spectra of HD\,220787, HD\,955, Hip063541, and Hip081166 for the nights of June 6 2005, June 7 2005, May 22 2012, and June 6 2012, respectively. Prior to scaling and removing telluric spectra from science spectra, we identified and removed absorption lines that were intrinsic to the stars. Stellar cubes were then used for the flux calibration. Namely, we opted to use HD\,146332 and HD\,220787 for June 6 2005, HD\,193933 for June 7 2005, Hip063541, Hip095404, and Hip098312 for May 22 2012, Hip081166 and Hip096923 for June 6 2012.

Our next task was to combine the flux-calibrated cubes of the different jitter positions and nights into a single mosaic. For this purpose, we aligned all images to the world coordinate system using the jitter offsets, occasionally correcting for small offsets pertinent to telescope pointing instabilities. We merged all available spectra, averaging flux values at any given wavelength and position on the sky. Median instead of average flux values were ascribed to pixels if the average value exceeded the corresponding median value by more than five times the standard deviation of the distribution. To obtain an ultra-deep (i.e., four-hour-long) exposure at the center of the mosaic, we combined all data after bringing the resolution of the H band cubes to that of the  H+K band cubes ($R$=1640). We trimmed outer parts of the mosaic with exposure times $\lesssim$1.5 hour, so that the signal-to-noise (S/N) ratios of the different regions are comparable. The final, fully reduced cube has a total FOV of 8.3$\arcsec $$\times$8.3\arcsec\ (or $\sim$2\,kpc$\times$2\,kpc). 

We identified spectral lines in the fully reduced cube, and removed the (stellar and AGN) continuum from each line within a range of 6000\kms. The continuum was calculated from neighboring spectral pixels along each spatial pixel. 

\section{Results}
\label{sec:results}

\subsection{Basic galaxy properties and regular gas motions}
\label{sec:results_basic}

The list of emission lines that we detected in the NIR spectrum of IC5063 comprises \paa , \brg , \hei\ 2.0587\um , \feii\ 1.6440 \um, \sivi\ 1.9634 \um , \sixi\ 1.9196 \um , \caviii\ 2.321  \um, \htwo\ (1-0) S(1), and \htwo\ (1-0) S(3). A kinematic analysis was feasible for these lines, which were bright enough to be little affected by neighboring sky/telluric lines.  The \htwo\ (1-0) S(3) line could be contaminated by \hei\ and/or \feii\ 1.954\um , $-$550\kms\ away \citep{vanzi08}. Because we can neither test this assumption nor evaluate the strength of these lines, we opt to show the \htwo\ (1-0) S(3) data, but to not use them in our kinematic analysis for velocities below $-$500\kms .
We detected numerous other emission lines in the spectra, but we did not create kinematic maps for them, either because their flux was low or because their profiles were substantially contaminated by sky/telluric lines. These lines comprise  \brd , Br10, Br11, \feii\ 1.5330\um , \htwo\ (1-0) S(0), S(2), S(5), S(7), and \htwo\ (2-1) S(1), S(3). The possibility that potential Br12 1.6412\um\ emission significantly affects the \feii\ 1.6440 \um\ emission (even though the Br13 and Br14 lines are undetected) is evaluated and rejected in Section~\ref{sec:results_jet_trail}. The \htwo (1-0) S(4) line was so contaminated by sky lines that its flux could not be recovered. Finally, we detected \htwo\ (1-0) Q(1), Q(2), and Q(3) emission, but the corresponding lines were discarded from our kinematic analysis because their restframe frequencies are only $\sim$1000\kms\ away and their fluxes are comparable: emission of ambient gas in one line could be mistaken for emission of outflowing gas in another line. In the rest of this work, when referring to \htwo\ lines, we will refer only to lines probing transitions from the first to the fundamental vibrational state unless otherwise noted.

\begin{figure*}[h!]
\centering
\includegraphics[height=22 cm, width=16cm ]{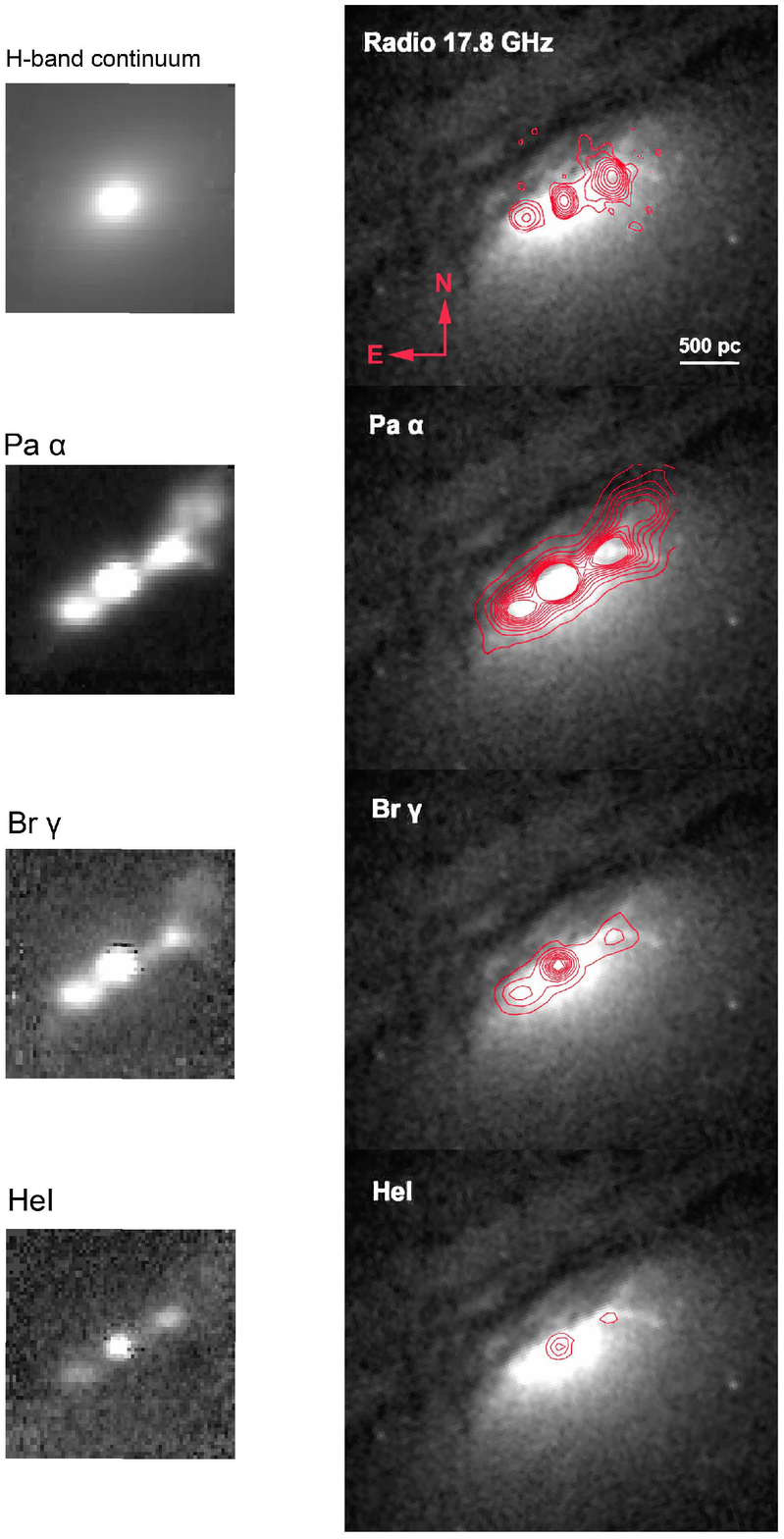} \\
\caption{ 
{\it Left:} images of the H-band continuum and of bright lines detected in the SINFONI data. The line images were created by collapsing continuum-free channels over the velocity range -500\kms $<$V$<$500\kms . {\it Right:} contours of these images (at 10$\sigma$ level steps going up to a maximum of 100$\sigma$) are plotted over a \hst\ WFPC2 547M image comprising \oiii\ 5007\AA\ emission \citep{schmitt03}. To indicate the radio jet projected position, contours of a 17.8 GHz image \citep{morganti07} are also overplotted in the top right panel.
} 
\label{fig:nir_contours}
\end{figure*}

The emission line maxima at the location of the nucleus indicate that IC5063 is at a redshift of 0.0116$\pm$0.0006. Collapsed images of the H-band continuum and of lines within $\pm$500 \kms\ from the systemic velocity are shown in Fig~\ref{fig:nir_contours}. The contours of these images, as well as the contours of a 17.8 GHz image \citep{morganti07}, are plotted over an optical (547M filter) image taken with WFPC2 onboard the \hst\ \citep{schmitt03}. The emission is spatially unresolved  in \sixi\ and \caviii , which probe the galaxy nucleus. The stellar CO(3-1) absorption seen near \caviii\ may be affecting the latter's distribution. In all other lines, the emission is resolved and follows the radio jet trail, like the \oiii\ 5007\AA\ emission in the WFPC2 image. \feii\ shows the overall most diffuse emission. Four distinct regions can be identified in the NIR images: the nucleus (at 20:52:02.40$-$57:04:07.68), the south radio lobe (1.8\arcsec\ or 450\,pc south east of nucleus, at 20:52:02.60$-$57:04:08.68), the north radio lobe (2.2\arcsec\ or 520\,pc north west of the nucleus, at 20:52:02.17$-$57:04:06.55), and the extended clump near the north-west edge of the NLR ionization cone (3.8\arcsec\ or 900\,pc away from the nucleus, at 20:52:02.06$-$57:04:05.05) that was previously seen by \citet{kulkarni98}.

\begin{figure*}[h!]
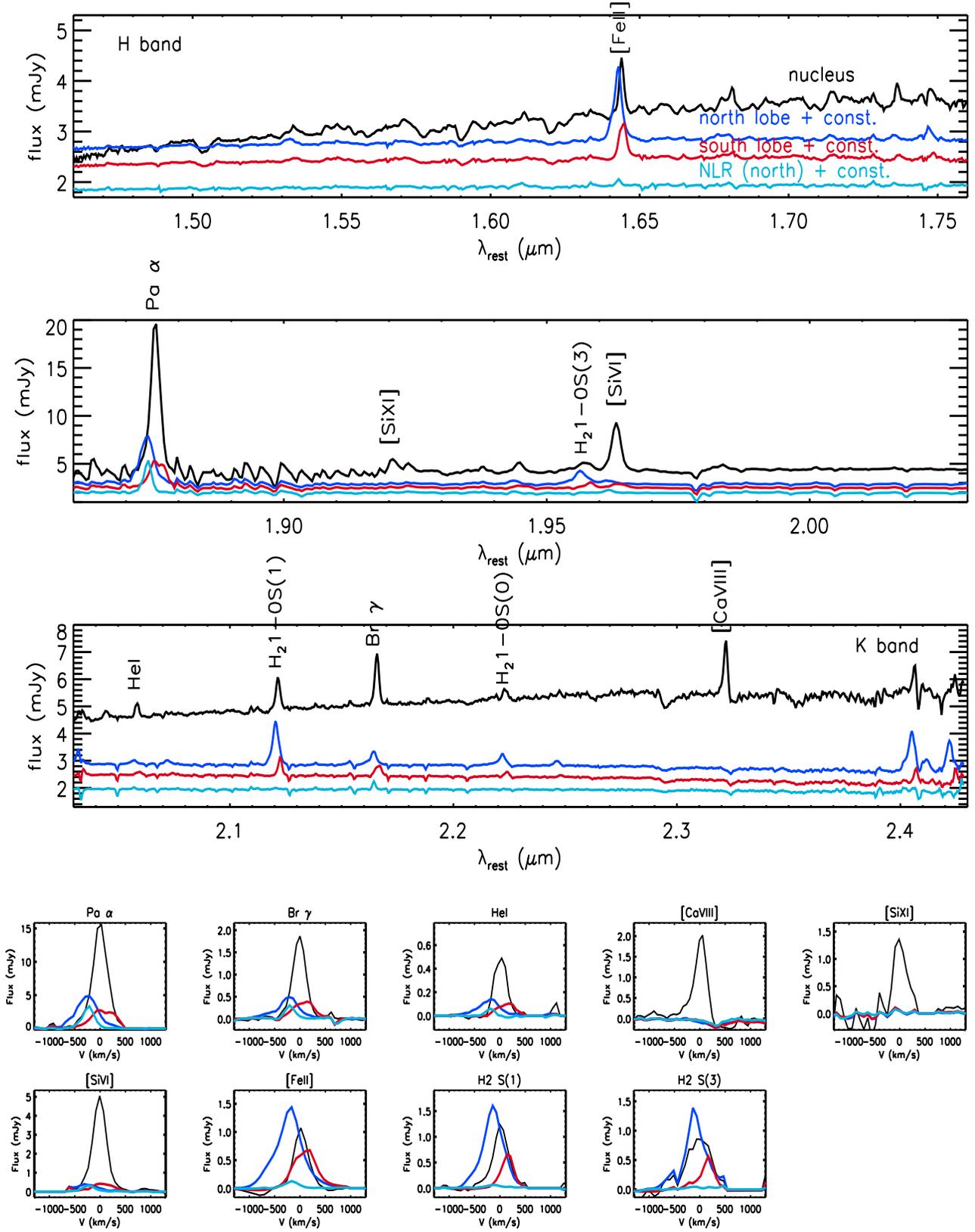

\centering
\includegraphics[height=16 cm, width=17 cm]{spectra_band.eps} 
\includegraphics[height=6 cm, width=17 cm]{spectral_lines.eps} 
\caption{ {\it Top:} rest-frame NIR spectra, extracted within a radius of 0.5$\arcsec$ from the nucleus, the north radio lobe, the south radio lobe, and the bright NLR clump of IC5063. A wavelength-independent constant of 
2 mJy, 1.5 mJy, and 1.5 mJy has been added to the spectra of the north lobe, of the south lobe and of the NLR clump, respectively, to facilitate their visual comparison. {\it Bottom:} continuum-free spectra of the same regions, 
zoomed in at spectral lines of interest. All panels are plotted using the same color scheme, i.e., black for the nucleus, blue for the north radio lobe, cyan for the NLR clump, and red for the south radio lobe. }
\label{fig:spectra}
\end{figure*}

Spectra of these four regions are shown in Fig.~\ref{fig:spectra}, integrated over a radius of 0.5\arcsec . The spectrum of the nucleus has a rising (from 2.4\,mJy to 5.6\,mJy) NIR continuum, characteristic of dust heated by the AGN. The highest number of detected lines is found in the nuclear region. \paa\ is the brightest of all, despite the strong atmospheric absorption window it is in. \caviii , \sivi , and \feii\ 1.6440 \um\ are among other bright lines. Contrarily, the region with the lowest number of detected lines is the NLR clump 3.8$\arcsec$ from the nucleus. Its spectrum has a much flatter and fainter continuum, of $\sim$0.4 mJy, as a result of the lower AGN and stellar emission. The slope and the intensity of the continuum at the north radio lobe and at the south radio lobe are intermediate of those at the nucleus and at the NLR clump. However, the \htwo\ and \feii\ emission is enhanced at the radio lobes - in particular at the north radio lobe, where these lines are brighter than at any other position in the central 2$\times$2 kpc$^2$ of IC5063. 

\begin{figure*}
\centering
\includegraphics[width=0.9\textwidth]{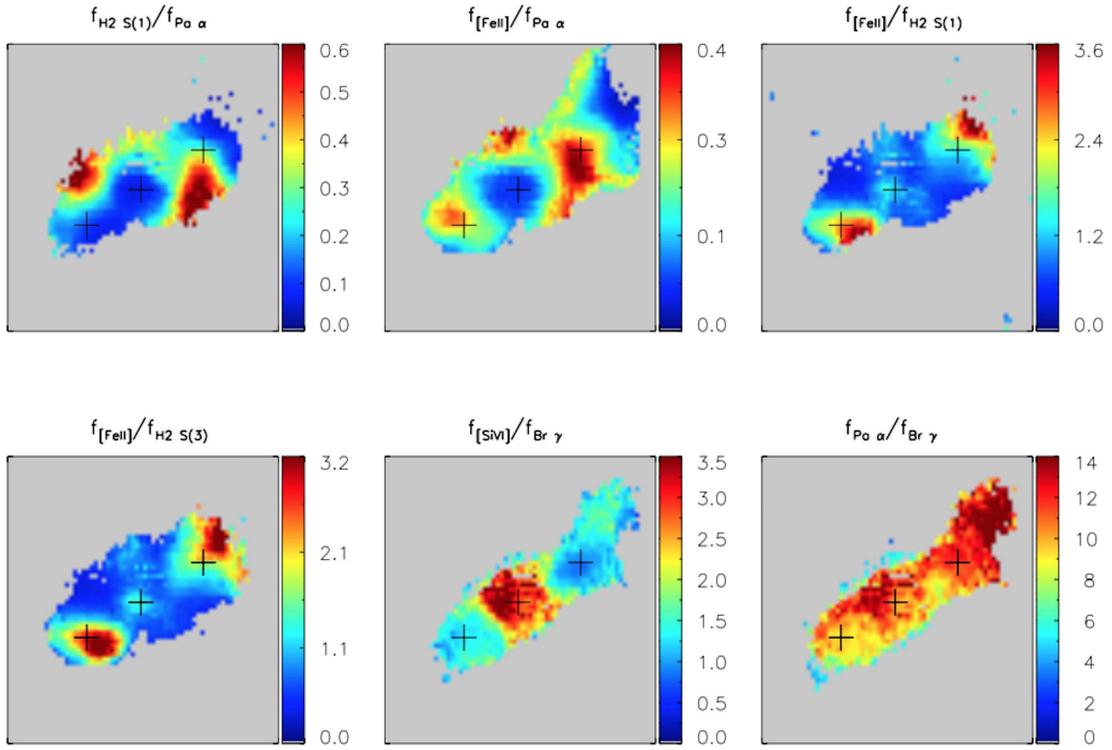}  
\caption{ Line ratio maps, created from the images presented in Fig.~\ref{fig:nir_contours} (and comprising the emission from gas with velocities $-$500\kms $<$V$<$500\kms ). Crosses mark the gas emission maxima at zero velocity at the nucleus and near the radio lobes.
}
\label{fig:emission_normalized_by_HI}
\end{figure*}

The flux ratio of \htwo\ (1-0) S(1) over \paa\ is 2-8 times lower at the nucleus than within 100\,pc from the north radio lobe, and $\lesssim$15 times lower further south of this region (Fig.~\ref{fig:emission_normalized_by_HI}).  Likewise, this ratio is up to 2 times lower at the nucleus than within 100\,pc from the south radio lobe, and $\lesssim$10 times lower further north of this region. The regions with emission excess from molecular gas, which could be star-forming or associated with shocked media \citep{ogle10,guillard12}, do not necessarily show emission excess from iron ions. The \feii /\paa\ flux ratio peaks mid-way between the radio lobes and the regions with \htwo\ emission excess, whereas the \feii /\htwo\ (1-0) S(1) and \feii /\htwo\ (1-0) S(3) flux ratios are maximum in regions where the \htwo\ (1-0) S(1)/\paa\ flux ratio is minimum. High fractions of ionized to molecular gas emission are seen in regions elongated perpendicularly to the jet propagation axis. Such regions are likely to contain shocked gas, because Fe requires the passage of shock fronts to be liberated from dust grains and massively deposited into a diffuse gas reservoir. The borders of the NLR are prominently outlined by ridges of high \feii /\paa\ flux ratio, either due to shocks or AGN illumination. The clump near the NLR north-west edge is bright in \sivi . The logarithm of the \sivi /\brg\ flux ratio in that region is in the range 0.1-0.3, which renders it compatible with shock and shock precursor ionization models, but also with AGN photoionization \citep{allen08,groves10,bicknell13}.

 The effects of obscuration are evaluated by means of comparison of the \paa /\brg\ ratio with the theoretically predicted value of 12 for case B recombination at 10$^4$K. In most regions, the ratio takes values between 12 and  9 (Fig.~\ref{fig:emission_normalized_by_HI}). These numbers respectively translate to optical depths of 0.0 to 2.2 in the H band, and 0.0 to 1.6 in the K band, assuming that the absorption follows the Milky Way R$_{\rm V}$=3.1 model of \citet{draine03}. Data reduction artefacts and differences in the gas excitation conditions can explain values outside this range. We find that there is little or no obscuration near the north radio lobe, near the nucleus, near the NLR most illuminated edge, and in front of the dust lane (north of the jet trail). Dust could be affecting the observed gas properties $\sim$300\,pc west of the nucleus and near the south radio lobe, but any such effects are not quantifiable. Regions with high obscuration and \htwo/\paa\ excess coincide in projection with large-scale structures, rich in molecular gas \citep{morganti15}.
 
To determine the kinematic properties of the gas in the center of IC5063, we fit a Gaussian function along the profile of each line in each spatial pixel. Only pixels with S/N$>$10 were used. The results of this computation are presented in Fig.~\ref{fig:fit_results}. The disk previously known from HI observations \citep{morganti98} is clearly seen in all of our tracers. It has a projected minor axis of 240($\pm$30)\,pc, calculated from the \hei\ map and from the hydrogen recombination line maps in regions displaying circular orbits. Its south-east side is moving away from us and its north-west side is coming toward us. On the south-east side, the average circular velocity \vcirc\ is 185$\pm$21\kms\ at 750\,pc from the nucleus, and 250$\pm$12\kms\ at 900\,pc from the nucleus. On the north-west side, the disk is rotating with \hbox{$-$157$\pm$42\kms }, \hbox{$-$190$\pm$44\kms }, and \hbox{$-$247$\pm$7\kms\ } at 800\,pc, 1\,kpc, and 1.25\,kpc from the nucleus, respectively. A confirmation of these numbers comes from the radio lobe spectra, where the jet-related shocks enhance the emission of the gas rotating with the disk. There, the shocked gas emits so prominently that disk-related spectral line displacements of order 200\kms\ are visible in the scale of the entire NIR spectrum (Fig.~\ref{fig:spectra}).  The resolution-corrected velocity dispersion of the gas, as measured from the same gas tracers as those used for the rotational velocity measurement, is 115$\pm$20\kms\ at the nucleus and 95$\pm$9\kms\ one kiloparsec north west of it. 

The examination of data cube slices around lines of interest reveals further details on the kinematics of the different gas components. The disk rotation can be clearly seen in Fig.~\ref{fig:kinematics}, where each panel shows an image with a velocity bin equal to two spectral pixels of the initial cube: the bulk of the emission is shifted from the west of the disk to the east of the disk with increasing velocity. The emission from the north-west branch of the NLR peaks at about $-$320\kms , and it can be seen for another $\pm$300\kms . Its counterpart on the other side of the galaxy, the south-east branch of the NLR, can be barely detected in the 157\kms\ and 315\kms\ \paa\ bins. Faint structures almost perpendicular to the disk major axis (at position angles between 25\degree\ and 50\degree ) can also be seen at zero velocity in \feii\ and \paa . One of them starts at the center of the galaxy and continues south west, whereas the other starts mid-way between the nucleus and the south radio lobe and continues north east. They are both linked to the large-scale spiral arms seen in CO by \citet{morganti15}.

\subsection{Gas near the radio lobes and along the jet trail: proof of its acceleration by the plasma} 
\label{sec:results_jet_trail}

\begin{figure*}[h!]
 \begin{center}
\includegraphics[width=0.82\textwidth, height=1.2\textwidth]{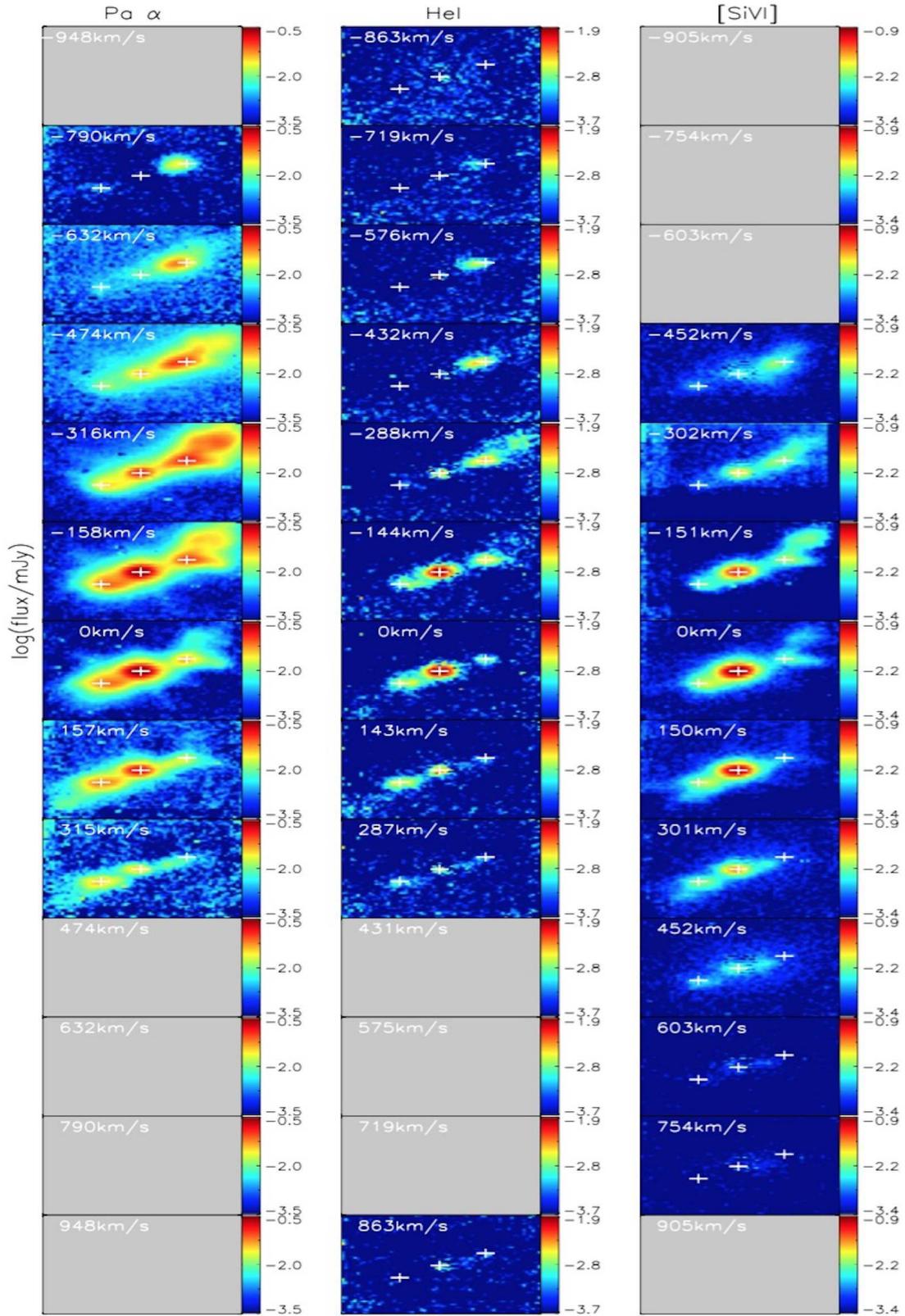} 
\caption{ Continuum-free slices of the SINFONI data, binned by a factor of two. Empty frames correspond to channels that have been masked because of the presence of 
sky/telluric residuals or neighboring ISM lines that could be confused with high-velocity wings. Crosses mark the gas emission maxima at zero velocity at the nucleus and near the radio lobes. }
\label{fig:kinematics}
\end{center}
\end{figure*}

\begin{figure*}
\centering
\begin{tabular}{l}
\includegraphics[width=17.8 cm]{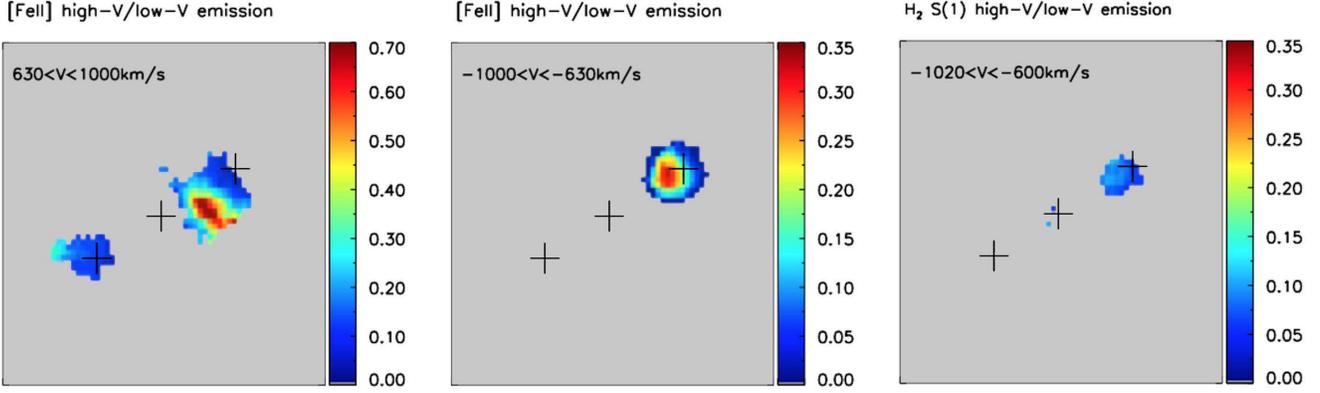} 
\end{tabular}
\caption{ Ratio of the outflowing gas emission over the ambient gas emission. The fraction is computed for velocity bins of 1\kms , because of the different velocity ranges spanned by the two components. 
Velocities in the range $-$250$<$V$<$250\kms\ were associated with the ambient medium. For the gas in the outflow, we used velocities $<$$-$600\kms or $>$600\kms\ (exceeding the disk circular velocity by more than three times the disk velocity dispersion). The outflow velocity range is noted in the upper left corner of each panel. Signal below three times the noise of the collapsed image has been masked.  At negative velocities, the maximum predicted Br12 contribution has been removed from the \feii\ image.}
\label{fig:fast_outflow_flux_ratios}
\end{figure*}

A striking result in Fig.~\ref{fig:kinematics} is the detection of high-velocity gas near the projected positions of the radio lobes, where the jet is thought to interact with dense ISM clouds. Near the north radio lobe, the high-velocity flow is seen for multiple species (H, He, Fe, Si, \htwo ) and for all gas phases (i.e., atomic ionized/neutral and molecular). \feii , for example, is detected over a continuous velocity range of 2000\kms\ (Fig.~\ref{fig:kinematics}). The \feii -emitting gas at $\sim$1000\kms\  (Fig.~\ref{fig:kinematics}) is moving against the disk rotation by $\sim$1200\kms . In the direction of the disk rotation, about half of the \feii\ emission comes from gas that is moving faster than the circular velocity along the line of sight (Fig.~\ref{fig:spectra}; bottom).  Despite the potential existence of Br12 1.6412\um\ emission \hbox{$-$510}\kms\ away, we confirmed that the detection of the highly blueshifted \feii\ wing is significant: under the assumption that the Br12 flux can be as high as the Br11 flux, we removed the stacked Br11 image at \hbox{$-$300}$\pm$180\kms\ from the stacked \feii\ image at \hbox{$-$810}$\pm$180\kms , and verified that the detection remains significant  (Fig.~\ref{fig:fast_outflow_flux_ratios}).  Gas that is moving faster than \vcirc\ by more than 600\kms\ is seen in several other lines, including \htwo\ (1-0) S(1).

Near the south radio lobe, the jet effects on the gas kinematics are only a little less pronounced than near the north radio lobe, even though the gas emission is lower by a factor of 2 (Fig.~\ref{fig:spectra}; lower panels). With \vcirc\ rising quickly from $\sim$100\kms\ to $\sim$200\kms\ near the south lobe (Fig.~\ref{fig:fit_results}), we again detect gas moving with velocities 600\kms\ above the disk circular velocity and 900\kms\ against it (Fig.~\ref{fig:kinematics}). This result, found for two sites, unambiguously shows that the irregular gas kinematics are due to jet-cloud collisions - not due to some coincidence (e.g., debris of a past merger; \citealt{colina91}; in the line of sight behind one of the lobes). 

\begin{figure*}[h!]
\begin{center}
\includegraphics[width=12.5 cm]{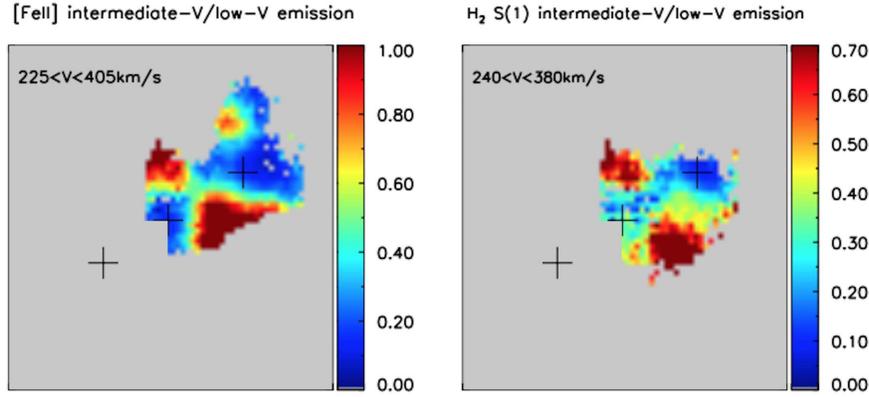} 
\caption{  As in Fig.~\ref{fig:fast_outflow_flux_ratios}, but now using the \feii\ emission at 315($\pm$90)\kms\ (left panel) and the \htwo\ (1-0) S(1) emission at 310($\pm$70)\kms\ (right panel) instead of in the fast outflow. Regions east of the nucleus have been masked, as they can be contaminated by the regular disk emission at these velocities.} 
\label{fig:slow_outflow_flux_ratios}
\end{center}
\end{figure*}

To study the outflow extent between the radio lobes, we created maps of the outflowing-to-ambient gas emission (in \feii\ and \htwo ). We did so for two velocity ranges of the outflowing gas. With the fast outflow used in Fig.~\ref{fig:fast_outflow_flux_ratios}, we associated gas moving faster than 600\kms\ (i.e., with a velocity exceeding \vcirc\ by more than three times the disk velocity dispersion). With the intermediate-velocity outflow used in Fig.~\ref{fig:slow_outflow_flux_ratios}, we associated gas moving slower than 600\kms , but against the disk rotation. To be conservative, we used the maximum measured circular velocity, 250\kms , and velocity dispersion, 115\kms\ in our definitions. By ambient, we refer to gas moving slower than \vcirc\ (in either direction). Fig.~\ref{fig:fast_outflow_flux_ratios} shows that the fast outflow extends along the radio jet trail from the vicinity of the nucleus to the north radio lobe. The intermediate-velocity outflow extends further away (Fig.~\ref{fig:slow_outflow_flux_ratios}; see also \citealt{morganti15} for a 200\kms\ outflow of cold gas probed by CO). 

Fig.~\ref{fig:fast_outflow_flux_ratios} further reveals that the highly blueshifted and the highly redshifted emission near the north radio lobe are not cospatial. The emission at 810($\pm$)180\kms\ peaks 240\,pc south east of the emission at $-$810($\pm$)180\kms\ (Fig.~\ref{fig:fast_outflow_flux_ratios}). 

To identify discrete outflow starting points, we used \hst\ FOC optical images of high spatial resolution ($\sim$10pc). The FOC image shown in Fig.~\ref{fig:outflow_extent} is the average of three individual exposures probing polarized light at 0, 60, and 120 degrees, respectively. Taken with the F502M filter, these images comprise \hbeta\ and \oiii\ emission.  Regions of interest are marked on Fig.~\ref{fig:outflow_extent}. With R1 and R2, we mark the beginning and the end of a 120\,pc-long filamentary structure, which is perpendicular to the jet trail. The peak of the highly redshifted \feii\ emission is near R2. The dark region west of R2 is where the H-band optical depth is $\sim$2, where the \htwo /\paa\ ratio is high (Fig.~\ref{fig:emission_normalized_by_HI}), and where a large-scale CO structure reaches the inner gas disk \citep{morganti15}. With R3, we mark the NLR base, i.e., the region where the ionization cone becomes visible. With R4, we mark the peak of the highly blueshifted \feii\ and \htwo\ emission. There, the emitting region in the FOC data is oriented north-south (i.e., with an inclination with respect to the jet trail). R1, R3, and R4, are respectively located 240, 320, and 400 pc away from the nucleus.

Upstream the NLR cone and north west of R4, more outflowing gas is detected by its \feii\ emission at 315$\pm$90\kms\ (or $\sim$550$\kms$ against \vcirc ; Fig.~\ref{fig:slow_outflow_flux_ratios}). A discrete starting point within this flow can be seen by the NLR clump identified by \citet{kulkarni98}. We mark it with R5 in Fig.~\ref{fig:outflow_extent}.

\subsection{Diffuse gas out of dynamical equilibrium}
\label{sec:results_diffuse}

\begin{figure*}[h!]
\centering  \includegraphics[width=0.85\textwidth]{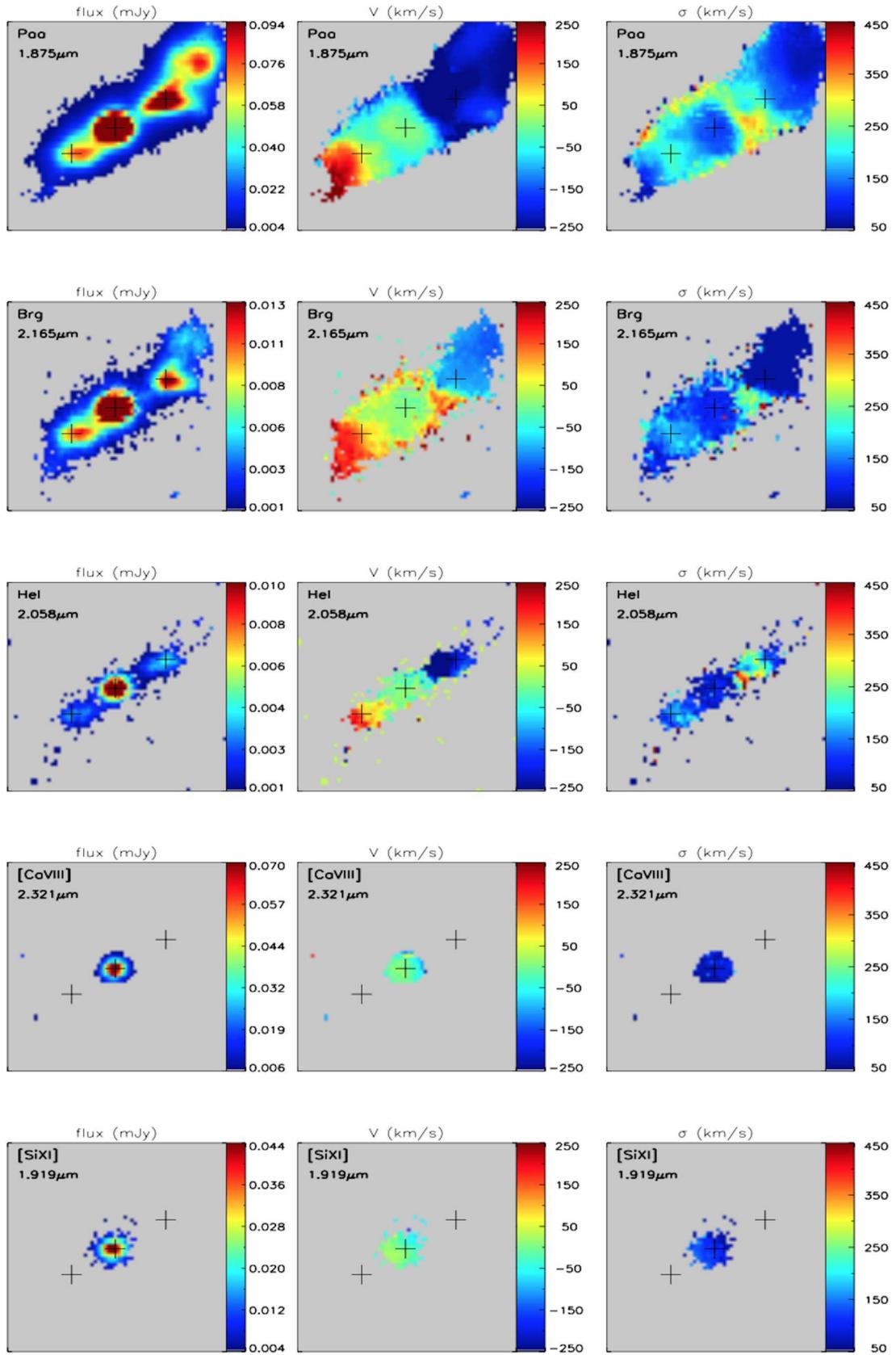} 
\caption{ Flux, velocity, and instrumental-broadening-corrected velocity dispersion of various gas tracers, as indicated by the best-fit solutions to the continuum-subtracted line spectra in each spatial pixel. Crosses mark the gas emission maxima at zero velocity at the nucleus and near the radio lobes. The dashed gray lines in the velocity field of \feii\ outline the biconical outflow perpendicular to the jet propagation axis.} 
\label{fig:fit_results}
\end{figure*}

The flow of accelerated gas is not just parallel to the jet trail. Fig.~\ref{fig:fast_outflow_flux_ratios} shows a diffuse outflow component that is oriented perpendicularly to the jet trail, and that spreads south west of R1. Its extent in either \feii\ or \htwo\ exceeds 300\,pc at high velocities and reaches $\sim$500\,pc at intermediate velocities (Figs.~\ref{fig:fast_outflow_flux_ratios},~\ref{fig:slow_outflow_flux_ratios}). The intermediate-velocity maps further reveal a similar diffuse outflow component spreading $\sim$400\,pc north east of R1 (Figs.~\ref{fig:slow_outflow_flux_ratios},~\ref{fig:outflow_extent}). The overall outflow has, thus, a biconical shape. Both its cones extend beyond the projected disk thickness (as derived from regions with circular orbits in the \hei\ and hydrogen recombination line maps). Conical shapes of gas moving against the disk rotation can also be seen in the \feii , \htwo\ (1-0) S(1), S(3), and \brg\ kinematic maps of Fig.~\ref{fig:fit_results}. 

The center of this biconical outflow is located 240\,pc north west of the galaxy nucleus. Its axis is perpendicular to the jet trail and parallel to the R1-R2 line in the FOC image, revealing an association between the outflow and the bright \hbeta\ and \oiii\ emission.  For \feii, the fraction of accelerated-to-ambient gas emission exceeds one near this line (Figs.~\ref{fig:fast_outflow_flux_ratios},~\ref{fig:slow_outflow_flux_ratios}). It is worth noting that this is not necessarily the highest intrinsic fraction, but the highest observed fraction due to ions scattered in our line of sight. The elongated region with the highest observed accelerated-to-ambient gas emission continues to the north as a ridge of high \feii\ velocity dispersion (Fig.~\ref{fig:fit_results}).  Along that ridge, the gas dispersion is boosted from 140($\pm$10)\kms\ to 290($\pm$20)\kms . On its west, the gas line-of-sight velocity goes down to \hbox{$-$350}($\pm20$)\kms\ in \feii , \paa , and \sivi . On its east, the iron ions are typically moving with 180-300\kms . 

\begin{figure*}[h!]
 \begin{center}
\includegraphics[width=0.8\textwidth ]{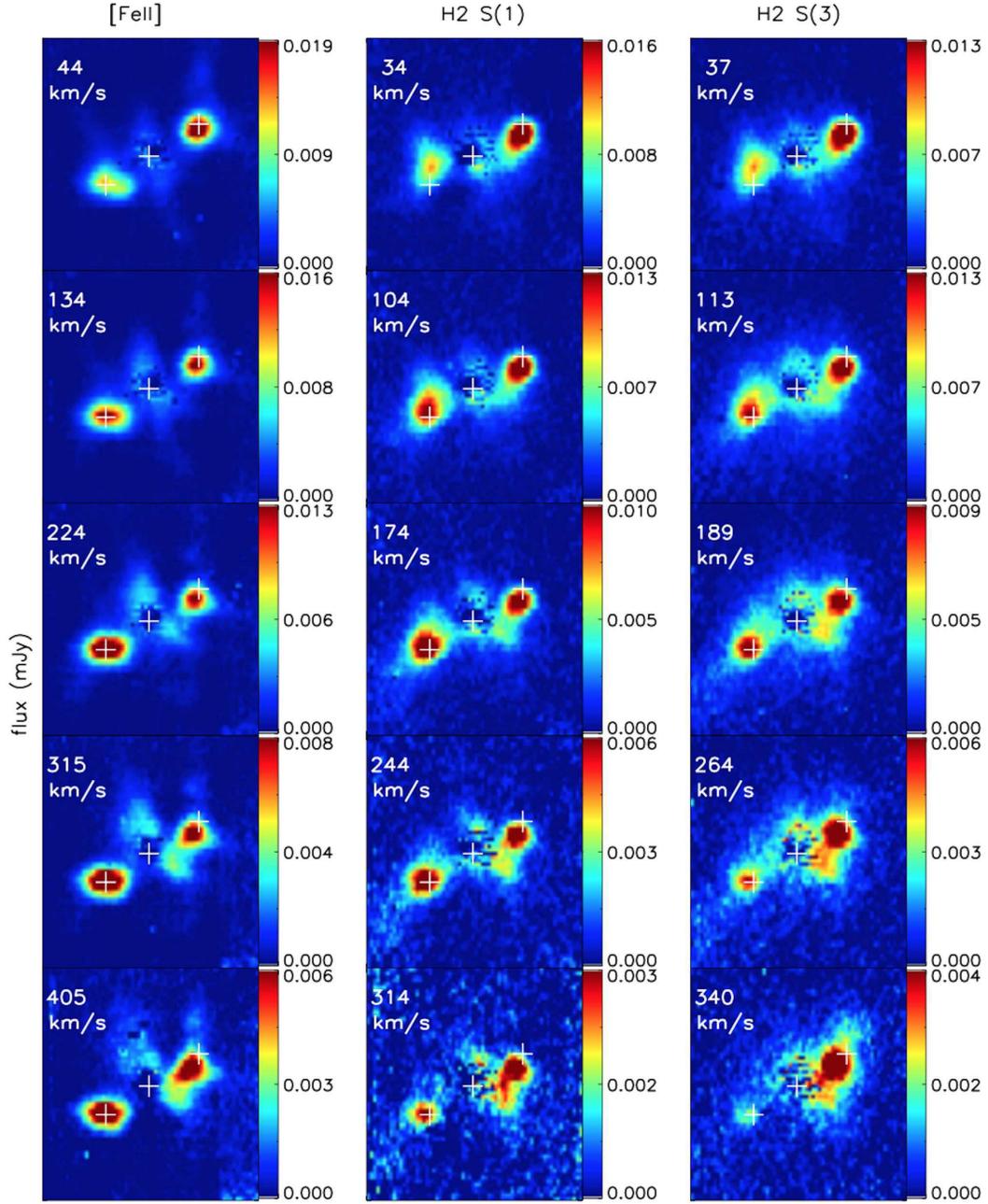} 
\caption{ Data frames showing the spatial progression of a faint structure from the nucleus toward the north radio lobe with increasing velocity. The disk emission, approximated by a \paa\ image at 0$\pm$75\kms\ that was normalized to the peak nuclear emission, has been removed from each frame to assist the visualization of the diffuse gas.} 
\label{fig:kinematics_faint_structure}
\end{center}
\end{figure*}

The data cubes provide enough information as to how the spatial distribution of the diffuse gas emission changes with velocity (Fig.~\ref{fig:kinematics_faint_structure}). In this Figure, the \feii , \htwo\ (1-0) S(1) and S(3) cubes are plotted at their finest velocity step, and the nuclear emission has been subtracted for visualization purposes. The diffuse emission from gas moving against the disk rotation is also visible prior to the nuclear flux removal (Fig.~\ref{fig:kinematics}). The inspection of the individual cube channels reveals that the gas emission spreads from the nucleus to the north lobe with increasing velocity. The region that the diffuse gas occupies unfolds almost symmetrically with respect to the jet trail.  At velocities $\gtrsim$240\kms , however, the  symmetry breaks as the emission drops sharply north of the trail, near the ridge of high \feii\ velocity dispersion. South of the trail, the emission continues, forming an arc that connects (in projection) the biconical outflow base with the north radio lobe. The arc is most pronounced for the molecular gas. An explanation for this progression in the outflowing gas spatial distribution as a function of velocity is given in the following Section. 

\begin{figure*}
 \begin{center}
\includegraphics[width=0.62\textwidth]{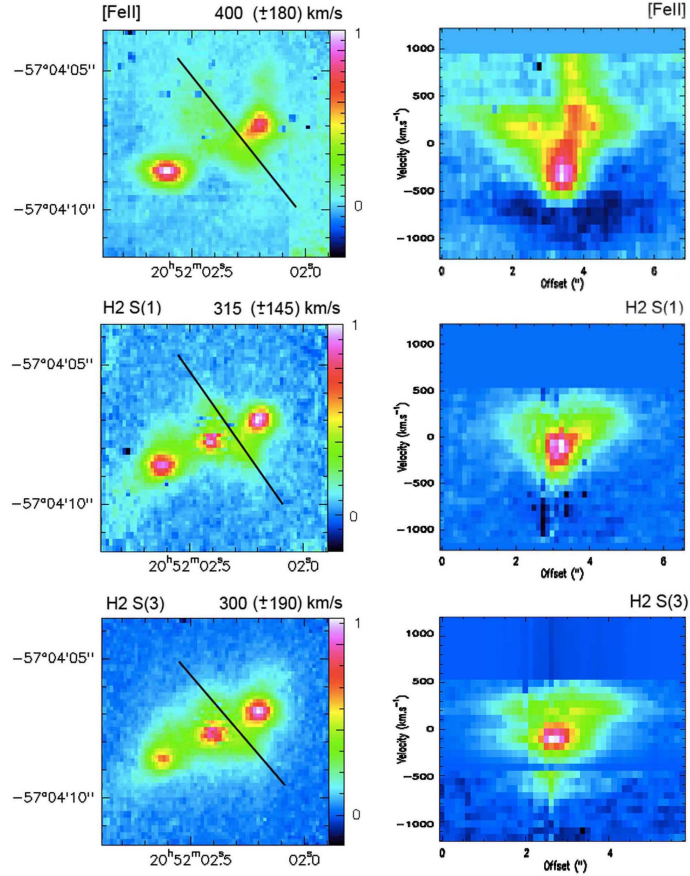} 
\caption{ Position-velocity diagram along a slice almost perpendicular to the jet propagation axis that passes from the \feii\ biconical outflow starting point. The offset along the slice increases from north east to south west and the flux is normalized to unity. The absorption at negative velocities in the \feii\ panel is real, due to the stellar CO(7-4) bandhead.  The absorption at $-$400\kms\ in the \htwo\ (1-0) S(3) panel is due to a sky line.} 
\label{fig:posang_perpendicular}
\end{center}
\end{figure*}

\begin{figure*}
 \begin{center}
\includegraphics[width=0.62\textwidth]{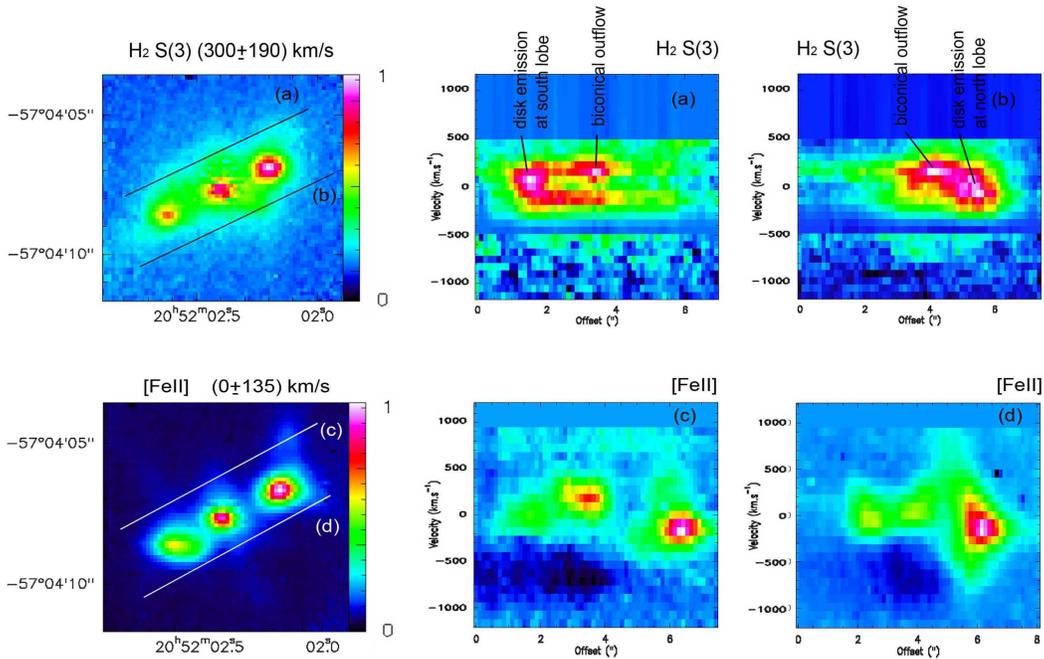} 
\caption{ Position-velocity diagram along slices almost parallel to the jet propagation axis. The offset along the slices increases
 from south east to north west and the flux in each panel  is normalized to unity. Letters are used to indicate which position-velocity diagrams correspond to which slices.
 }
\label{fig:posang_parallel}
\end{center}
\end{figure*}

Position-velocity diagrams along position angles pertinent to the biconical outflow are shown in Figs.~\ref{fig:posang_perpendicular} and ~\ref{fig:posang_parallel}.   In Fig.~\ref{fig:posang_perpendicular}, we examine a slice that is nearly perpendicular to the jet axis and that passes from the starting point of the biconical outflow. All along 800\,pc, gas is continuously detected  with a velocity peak of $\sim$ 200\kms\ for all of \feii , \htwo\ (1-0) S(1) and S(3). For \feii , we see that the emission spreads out to 1.2\,kpc, and that the gas attains a sharp terminal velocity of 400\kms . When reading this position-angle diagram from the jet propagation axis (i.e., from an offset of 3\arcsec ) to outer regions (i.e., to offsets of 0\arcsec\ or 6\arcsec), we find that the diffuse gas velocity increases with increasing distance. The velocity gradient is smooth, and close to symmetric in the two sides of the jet. 

In Fig.~\ref{fig:posang_parallel}, we examine slices nearly parallel to the jet axis. Along the northern border of the biconical outflow, which runs parallel to the jet propagation axis at a distance of $\sim$350\,pc, there is intermittent \feii\ emission at 900\kms\ along both sides  of the disk (Fig.~\ref{fig:posang_parallel}). This indicates that the ionized gas wind generated north west of the nucleus is strong enough to dominate the \feii\ emission even north east of the nucleus (see also Fig.~\ref{fig:fit_results}). This is not true for the warm molecular gas, even though \htwo\ (1-0) S(3) emission is also seen along 1.3\,kpc in both sides of the disk (at about $-$100\kms ). West of the nucleus, part of this emission is coming from the disk. East of the nucleus, the emission is blueshifted by 260\kms\ with respect to the disk rotational velocity.  It is, thus, likely due to shocks generated east of the nucleus (Figs.~\ref{fig:fit_results},~\ref{fig:posang_parallel}). The region with negative velocities north east of the nucleus can also be seen in the \htwo\ (1-0) S(1) and S(3) velocity fields (Fig.~\ref{fig:fit_results}). This is another region where gas flows in both line of sight directions, as \feii\ emission is also detected there at $\sim$900\kms .  Diffuse outflowing gas is, thus, also present on the east side of the disk. Its emission is, however, too weak to reveal changes in the gas spatial distribution as a function of velocity.

Notably, the (more tenuous) ionized gas attains higher velocities than the (denser) molecular gas at fixed positions (Figs.~\ref{fig:kinematics},~\ref{fig:kinematics_faint_structure},~\ref{fig:posang_perpendicular},~\ref{fig:posang_parallel}). This result is in agreement with multi-phase outflow acceleration models \citep{hopkins10, wagner11, zubovas14}.

\section{Discussion}
\label{sec:discussion}

\subsection{On outflow properties pertinent to galaxy evolution}

\begin{figure*}
\centering
\includegraphics[width=0.72\textwidth ,height=5.8cm]{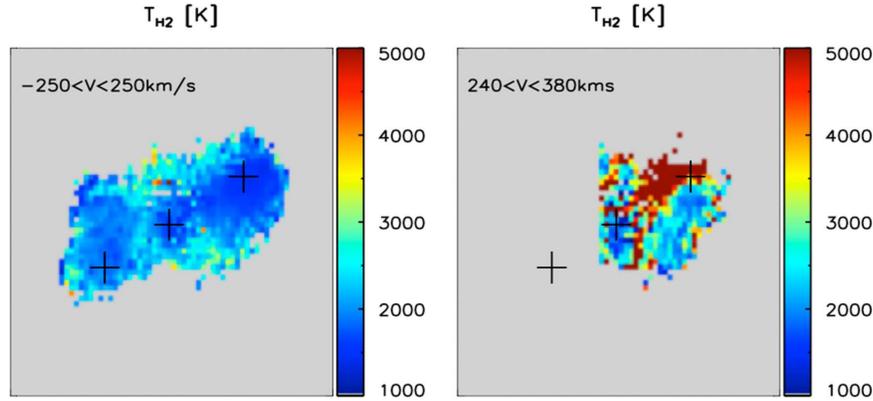} 
\caption{ Maps of the warm molecular gas excitation temperature in the ambient medium (left panel), and in the range 240\kms\ to 380\kms\ comprising diffuse gas that moves against the disk (right panel). The maps were calculated under the assumption of gas in LTE. In the right panel, the excitation temperature of the gas in the dark red pixels either exceeded 5000K, or it could not be computed under the assumption of thermalized gas. Emission from the disk could be contributing to the flux of both lines. For this reason, the side of the disk moving away from us has been masked. }
\label{fig:H2_temperature}
\end{figure*}

A question pertinent to galaxy evolution studies that NIR data help us to address is whether the jet heats up the molecular gas it accelerates. To evaluate this possibility, we compared the gas excitation temperature \tex\ in the ambient medium and in the intermediate-velocity diffuse outflow\footnote{This question cannot be evaluated for the fast outflow due to the potential blending of the \htwo\ (1-0) S(3) line with \feii\ or \hei\ lines at $-$550\kms .}.
For this purpose, we created temperature maps of the warm \htwo\ using the (1-0) S(1) and S(3) lines, assuming that the gas can be described by a single temperature component in local thermodynamic equilibrium (LTE). The results are shown in Fig.~\ref{fig:H2_temperature}. The ambient gas excitation temperature at the north radio lobe is 1500$\pm$200\,K  \citep[see also ][]{tadhunter14}. In the rest of the disk, it is 2000$\pm$200\,K. At the edges of the disk, near the regions with excess of molecular-to-atomic-gas emission (Fig.~\ref{fig:emission_normalized_by_HI}), it goes up to 2700$\pm$500\,K. The low temperature of the ambient gas at the north radio lobe could be indicative of a region that is denser and more self-shielded than its surroundings, which can be described by several components of subthermally excited gas. 

For the diffuse accelerated gas at 310$\pm$70\kms , \tex\ varies significantly from one region of the map to another. In the south of the jet trail, \tex\ is within the range of values found for the ambient medium. However, no \tex\ solution can be found for many pixels north of the jet trail:
north of the regions R3, R4, and the north radio lobe (where diffuse 17.8\,GHz emission exists; Fig.~\ref{fig:outflow_extent}), the (1-0) S(3)/S(1) flux ratio is above the saturation limit (2.04) for a population that is described by a Boltzmann distribution and that decays spontaneously (Fig.~\ref{fig:H2_temperature}). This result implies that the molecular gas is in part non-thermally excited and that radiative decay in the form of fluorescence is plausible. The non-thermal source of excitation could be X-rays or UV light coming, for example, from the jet-related high-velocity shock wave or from the AGN radiation \citep{lepp83,black87,draine90,mouri94}. This result does not rule out a temporary temperature increase of the gas that got exposed to the non-thermal source of excitation: radiative cascade and heating are often coupled processes \citep[e.g.,][]{mouri94}. Evidence of heating comes from the work of \citet{young07}, who found that the dust is warmer at the north radio lobe than at south radio lobe.
                                         
\begin{figure*}
\centering
\includegraphics[width=0.72\textwidth ,height=5.8cm]{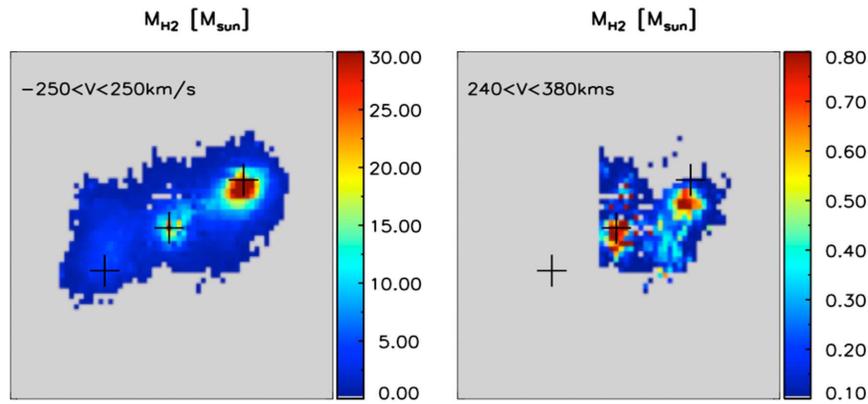} 
\caption{ Maps of the warm molecular gas mass in the ambient medium (left panel), and in the range 240\kms\ to 380\kms\ comprising diffuse gas  that moves against the disk (right panel). The maps are calculated for the excitation temperature maps of Fig.~\ref{fig:H2_temperature}. In the right panel, emission from the disk could be contributing to the measured \htwo\ mass.}
\label{fig:H2_mass}
\end{figure*}

 The \htwo\ mass distribution that we can create when taking the \tex\ map at face value and assuming an ortho-to-para ratio of 3 is shown in Fig.~\ref{fig:H2_mass}. The total ambient gas mass is 3500\msun\ (of which 990\msun , 460\msun , and 150\msun\ are found within a radius of 0.5\arcsec\ from the north lobe, the nucleus, and the south lobe, respectively). For the gas at 310$\pm$70\kms , the total mass is 170\msun\ (of which 30\msun\ and 80\msun\ are within 0.5\arcsec\ from the north lobe and the nucleus, respectively). The remaining 60\msun\ come from the diffuse medium. Fig.~\ref{fig:H2_mass} demonstrates that the maximum mass of the diffuse molecular gas (that can be detected from our line of sight) follows a filamentary distribution near the biconical outflow axis. The filamentary structure carries 35\msun . For comparison, the mass of the molecular gas in the fast outflow at $-$810($\pm$)210\kms\ would be 25\msun , if all of the \htwo\ (1-0) S(1) emission in it was due to thermalized gas at \tex =2000\,K. The total mass in the outflow at all velocities and regions is, thus, unlikely to exceed some hundreds of solar masses \citep[see also ][]{tadhunter14}. We, thus, conclude that the contrast between the mass carried in the warm phase and the mass carried in the cold phase \citep[2.3$\times$10$^7$-1.3$\times$10$^8$\msun ; ][]{morganti13} remains stark. If the acceleration of the molecular gas happens through its partial dissipation and reformation as proposed by \citet{hopkins10}, then this result is a proof that microturbulence \citep{nesvadba11} ceases rapidly enough for cooling and potentially for star formation to occur during the outflow propagation timescale.

\subsection{The jet and its cocoon as drivers of the gas kinematics}

The bottom line of the results presented in Section~\ref{sec:results} is that numerous gas components with irregular kinematics are detected in the central square kiloparsec of IC5063. Which mechanism drives the observed gas motions? In this Section, we argue that the propagation and the deflection of the radio jet plasma does so.

\begin{figure*}
\centering
\includegraphics[width=16 cm, height=4.7 cm]{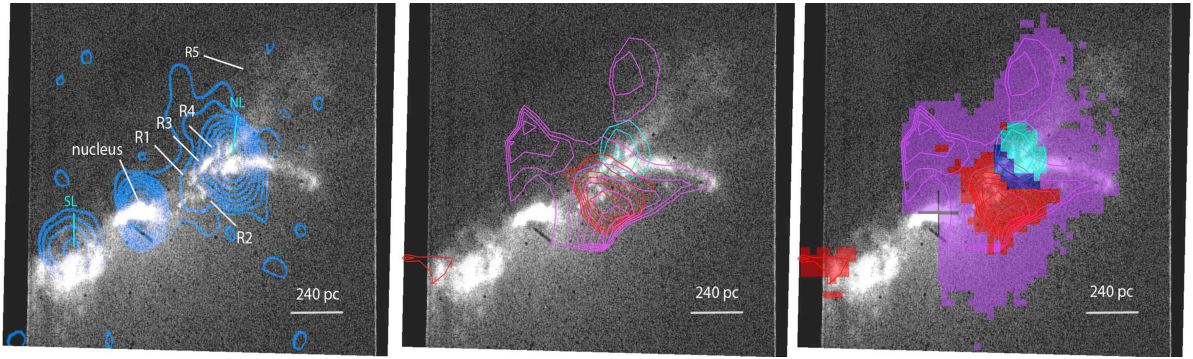} 
\caption{ Regions of interest with respect to the radio jet trail (left panel) and the outflow (middle and right panels) are plotted over a high angular resolution \hst\ FOC image of IC5063. The FOC F502M filter was used, which comprises
\hbeta\ and \oiii\  emission. 17.8\,GHz contours from \citet{morganti07} are plotted in blue in the left panel. The north lobe and south lobe are marked with NL and SL. Contours of the outflowing-to-ambient \feii\ emission (from Figs.~\ref{fig:fast_outflow_flux_ratios}, ~\ref{fig:slow_outflow_flux_ratios}) are plotted in the middle panel, revealing regions with a high fraction of outflowing Fe ions. Cyan, red, and magenta curves correspond to highly blueshifted, highly redshifted, and intermediately redshifted emission, respectively. The contours are at steps of 0.1 starting from 0.1 for the fast outflow, and at steps of 0.25 starting from 0.25 for the intermediate-velocity outflow. In the right panel, we plot a map of the outflow in either gas phase (atomic or molecular), using pixels where $>$10\% of the emission comes from outflowing gas. The flow extent at intermediate velocities is shown in magenta. The fast outflow extent is shown in red for the region with gas flowing away from us, in blue for the region with gas flowing toward us, and in purple for the region with gas flowing in both directions.
}
\label{fig:outflow_extent}
\end{figure*}

The diffuse gas geometric characteristics are not in line with an AGN-related or a starburst-related mechanism of acceleration. The biconical outflow in IC5063 cannot be associated with a bubble driven by the AGN radiation (as in NGC\,4258; \citealt{jimenez10}), because it is not centered at the nucleus. The gas kinematics are also incompatible with those of an expanding bubble: we observe no spherical symmetry in the velocity gradient of the accelerated Fe ions. No ultra-fast outflow \citep[as in][]{wagner13} has been detected in X-ray spectra of IC5063 \citep{tombesi14}, and no \caviii\ or \sixi\ outflow is seen in the SINFONI data. A nuclear starburst is ruled out for the same symmetry reasons as those applicable to the AGN. An obscured starburst located near region R1 of Fig.~\ref{fig:outflow_extent} is also unlikely. A biconical outflow similar to that in, e.g., M82 \citep{walter02,veilleux09} would mainly show redshifted emission in the northern hemisphere of the galaxy and blueshifted emission in the southern hemisphere of the galaxy. The SINFONI data indicate a symmetry of the redshifted emission in the north and south hemispheres, and an asymmetry of the redshifted vs. blueshifted emission to the west and east of region R3. 

Contrarily to an AGN or starburst driven wind, numerous findings support the jet-ISM interaction: the gas emission is brighter north west than south east of the nucleus (like the plasma emission itself), the non-thermally excited gas lies parallel to the jet trail,  outflows are initiated in at least four discrete regions along the jet trail, the highest gas line-of-sight velocities ($<$$-$600\kms\ or $>$600\kms ) are attained near the proposed jet-ISM interaction points, and the gas emission can be blueshifted/redshifted in the same hemisphere. 

As the plasma propagates, bow shocks push the surrounding tenuous gas and form an expanding cocoon. On its turn, the cocoon applies ram pressure to clouds that it encounters as it expands. The emission from the latter clouds can be either blueshifted or redshifted on the same side/hemisphere of the galaxy, depending on the positioning of the jet and the gas with respect to us. If the jet travels between us and background clouds, then we see redshifted emission from the background clouds that are being pushed further away from us by the cocoon. If foreground clouds exist between the jet and us, then we see blueshifted emission from the foreground clouds that are being pushed closer to us by the cocoon \citep[see also][]{garciaburillo14}. In regions where the jet plasma directly impacts clouds, then both blueshifted and redshifted emission from the scattered material can be detected. If the jet propagates through a clumpy disk, the plasma can get deflected on several dense regions and the cocoon can push several clouds. Then, the cloud distribution is most important for determining how many outflows will be initiated and in which directions they will be headed  \citep{wagner11}. 

\begin{figure*}
\centering
\includegraphics[width=0.48\textwidth ]{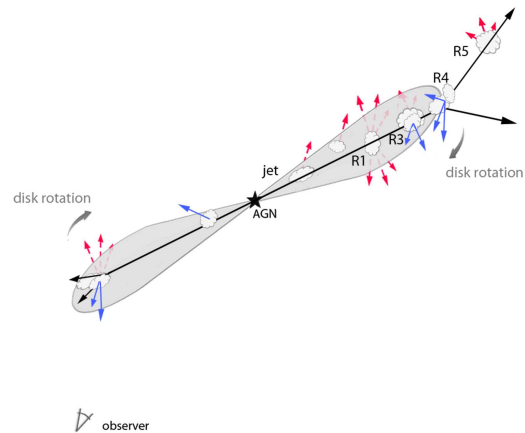} 
\caption{ Schematic representation of the jet-ISM interaction. Solid black lines represent the radio jet. The gray area depicts the cocoon in the plane of the disk.
Clouds located behind the radio jet with respect to the observer are pushed further back by the jet plasma/cocoon. Their outflow is characterized by redshifted emission (red arrows) in either galactic hemisphere.
Inversely, clouds located in front of the radio jet are pushed nearer to the observer (blue arrows). The region nomenclature follows that of Fig.~\ref{fig:outflow_extent}.
}
\label{fig:jet_schematic}
\end{figure*}

One schematic representation of the jet-ISM geometry explaining all observed outflows in IC5063 is presented in Fig.~\ref{fig:jet_schematic}. It follows the cloud naming convention of Fig.~\ref{fig:outflow_extent}, under the assumption that the large-scale \hbeta\ and \oiii\ emission in the FOC image is associated with the jet trail, as advocated by \citet{morganti07}. Fig.~\ref{fig:jet_schematic} shows a jet propagating through a clumpy medium and its cocoon on the disk plane. North west of the nucleus, most clouds  are located behind the jet, whereas south east of the nucleus,  most clouds are located in front of the jet (with respect to us). This is in line with several results:  with the small inclination of the jet with respect to the disk plane (Fig.~\ref{fig:outflow_extent}; \citealt{morganti07}) making the north west radio lobe visible below the dust lane in the southern galaxy hemisphere, with the NIR obscuration being higher at the south east radio lobe than at the north west radio lobe (Fig.~\ref{fig:emission_normalized_by_HI}), and with the gas emission in the FOC image being diffuse south east of the nucleus - possibly due to high ram pressure (Fig.~\ref{fig:outflow_extent}). 

In this picture, the jet and its cocoon drive a mostly redshifted outflow north west of the nucleus and a mostly blueshifted outflow south east of the nucleus, clearing out some of the gas in the disk. North west of the nucleus, consecutive major jet-ISM interactions are occurring, starting at the base of the biconical outflow (near R1; Fig.~\ref{fig:jet_schematic}). There, the passage of the plasma by the near side of a dense ISM structure, as that seen in the ALMA CO(2-1) data \citep{morganti15}, causes the redshifted biconical emission. Part of the plasma is likely to be scattered to the south of the disk plane, as indicated by the weak 17.8\,GHz emission, which extends along the southern biconical outflow axis and which ends where the fast outflow ends \citep{morganti07}. Then, near the north radio lobe, in region R4, an interaction between the jet and the far side of another gas clump generates the outflow coming toward us (Fig.~\ref{fig:jet_schematic}). This clump, oriented north south, could have been shaped by plasma that got deflected at the north radio lobe and that drilled a channel on its way out of the disk (Fig.~\ref{fig:outflow_extent}). Again, the 17.8\,GHz image of \citet{morganti07} shows diffuse radio emission north of region R4 (starting north of R3 and ending near R5). Between regions R1 and R4, at the NLR base (R3), we see either the spatial overlap of the outflows at R1 and R4, or another jet-ISM interaction that scatters gas in many directions. Further up the NLR cone, in region R5, the jet again pushes away from us the near side of a cloud. On the other side of the nucleus, near the south radio lobe, scattered gas is accelerated in both lines of sight near the tip of the radio jet (Figs.~\ref{fig:outflow_extent},~\ref{fig:jet_schematic}). 

The progression in the location of the diffuse gas emission with velocity can be explained by the jet propagation in more than one way. We could be observing tenuous ISM material that got accumulated and accelerated before impacting the dense structure at R1, material that got scattered from that dense structure, or a combination of the two. In the first case, we observe the motions of a cocoon-driven outflow that gets mass loaded as it moves from the nucleus to the outer parts of the galaxy. The velocity of this outflow will be higher near the jet cocoon (i.e., closer to the dense structure at the biconical outflow starting point). Once at the dense structure, the mass-loaded outflow will appear deflected in two directions perpendicular to the initial trajectory because of line of sight effects. In the second case, we observe an outflow that got mainly mass loaded after the impact of the jet upon the dense structure, when gas got bounced to different directions. The gas scattered close to our line of sight appears to be moving faster and to be located closer to the scattering surface, whereas the gas scattered nearly perpendicular to our line of sight appears to be moving slower and to be closer to the nucleus.  In either way, scattering of the plasma links the gas entrainment along the jet trail and perpendicular to the jet trail, as well as in small and large scales. The high-velocity-dispersion regions near R1 to R4 (e.g., the ridge north east of R1 in the \feii\ kinematic map) are indicative of scattering. A spatial overlap of two accelerated/scattered gas components in projection or in volume (i.e., at the interface between two flows) can add to the observed dispersion.

The diffuse outflowing gas that we detected within the central square kiloparsec of IC5063 reaches hundreds of parsec away from visible jet breaking points. Its total north-east to south-west extent exceeds 700\,pc. In the south of the jet trail, the outflowing gas emission fills the entire volume between the galaxy minor axis and the NLR south-west branch (Fig.~\ref{fig:outflow_extent}). Likewise, in the north of the jet trail, the regions with an intermediate-velocity outflow (Figs.~\ref{fig:fit_results}, ~\ref{fig:outflow_extent}) together with the region of oversaturated \htwo\ (1-0) S(3)/S(1) emission (Fig.~\ref{fig:H2_temperature}), again fill the volume between the galaxy minor axis and the NLR north-west branch. This emission, thus, spreads for 700-900\,pc along its two south-east to north-west borders.  The outflow detection in an area of order 1 kpc$^2$ emphasises the role of jets in affecting the reservoir for star formation. Notably, the simulations of \citet{wagner11} mimicking the propagation of a jet in a clumpy disk, predicted that the turbulence would be so wide-spread. 

The distance out to which we detect the diffuse gas depends on the flow travel time or on the presence of a physical obstacle such as star-forming regions (with excess \htwo /\paa\ emission; Fig.~\ref{fig:emission_normalized_by_HI}), in addition to the depth of the observations. The gas travel time gives us a limit on the timescales of the (intermittent or not) relativistic plasma ejection. If the Fe ions were moving at an average projected velocity of 900\kms , then 4$\times$10$^5$ years would be required for them to travel from the disk to 350\,pc north of it. For the molecular gas moving at $-$100\kms , the timescale would be an order of magnitude higher, but it would be more uncertain: the fraction of molecular gas that forms in the flow from dense atomic gas and the pertinent timescales for its formation are unknown, rendering the use of a constant, low velocity unclear.

\section{Summary}
\label{sec:summary}
 
We used SINFONI data of IC5063, a nearby elliptical galaxy in which a radio jet propagates through a disk, to look for unambiguous observational evidence of the jet-ISM interaction and to study its imprints on the ISM. Multiple pieces of evidence were found, which can be summarized by the detection of high-velocity gas in (at least) four regions along/near the jet trail, and of diffuse accelerated gas that is associated with the jet cocoon.  Main results are as follows.

\begin{itemize}

\item
Emission is seen all along the spatially resolved jet trail in numerous NIR lines, including molecular and atomic (neutral/ionized) gas tracers. The brightest \htwo\ and \feii\ emission is located at the projected position of the north radio lobe. The south radio lobe and a clump in the NLR north-west branch are clearly identified.  Diffuse atomic and molecular gas emission is also seen between the nucleus and the north radio lobe.
\item
At the north radio lobe, we detect gas moving with $>$700\kms\ above the disk circular velocity and 1200\kms\ against it. Likewise, at the south radio lobe, we detect gas moving with 600\kms\ above the disk circular velocity and 900\kms\ against it. The derivation of this result for both sites unambiguously proves that the gas has been accelerated by a collision of the jet plasma with ISM clouds. 

\item
We also detect an outflow bright in \feii\ emission at 315\kms\ that starts near the north west branch of the NLR, and another outflow bright in \feii\ and \htwo\ emission that starts 240\,pc north west of the nucleus and that is associated with a filamentary structure perpendicular to the large-scale jet trail. This structure, which is characterized by strong \hbeta\ and \oiii\ emission and surrounded by extended 17.8\,GHz emission, could be due to deflected plasma. The newly detected outflow near this structure has a biconical shape. It entrains iron ions and \htwo\ molecules moving as fast as $-$810$\pm$180\kms\ and 310$\pm$70\kms , respectively. Along its southern axis, it shows the highest mass load in the line of sight, and higher \feii\ emission than in the ambient medium. 
 
 \item 
The spatial distribution of the diffuse outflowing gas changes as a function of velocity from  0\kms\ to 480\kms . More specifically,  the \feii , \htwo\ (1-0) S(1) and S(3) emission unfolds parallel to the jet propagation axis and extends further away from the nucleus with increasing velocity. Likewise, the velocity of the diffuse gas smoothly increases with increasing distance from the biconical outflow starting point both to the north east and to the south west of the jet trail. Again, plasma propagation and scattering explain the observed gas kinematic behaviour.

\item
Gas is driven out of dynamical equilibrium and dispersed over an area of order 1kpc$^2$ by the jet and its associated cocoon. In the west side of the disk, outflowing gas is detected in either phase (atomic/molecular) or direction for more than 700\,pc along the jet propagation axis. Parallel to this axis, but north and south of it, the emission from gas that is kinematically or excitationally affected by the jet fills the entire volume between the galaxy minor axis and the NLR. This emission, thus, spreads for $\sim$1\,kpc. The total north-east to south-west extent of the diffuse outflowing gas also exceeds 700\,pc.  Diffuse gas exists in the east side of the disk too. Its extent cannot be determined, however, because the diffuse emission from accelerated gas is weak there, like the radio emission.

\item
The highest line-of-sight velocities are attained near the areas where the plasma gets scattered. In the flow, the atomic gas typically reaches higher velocities than the molecular gas at fixed positions.

\item
The computation of an excitation temperature from the \htwo\ (1-0) S(1) and S(3) line fluxes, under the assumption that the gas is thermalized, fails for gas in the diffuse outflow near the north radio lobe.  This result implies that the gas is in part non-thermally excited and that radiative decay in the form of fluorescence is plausible. It doesn't, nonetheless, rule out gas heating.

\end{itemize}

\begin{acknowledgements}
K. D. acknowledges support by the European Commission through a Marie Curie Intra-European Fellowship 
(PIEF-GA-2013-627195;  'BHs SHAPING GALAXIES') awarded under the Seventh Framework Programme  
and by the European Research Council through participation to the Advanced Grant Program 267399 'Momentum' 
(PI Combes). The Maryland Summer Scholars Program and the Alumni Association of the College of Computer, Mathematical, 
\& Natural Sciences of the University of Maryland supported this project via internships awarded to A. C. B. in the summer of 2013.
\end{acknowledgements}

{}

\begin{figure*}[h!]
\centering
\includegraphics[height=22 cm, width=16cm ]{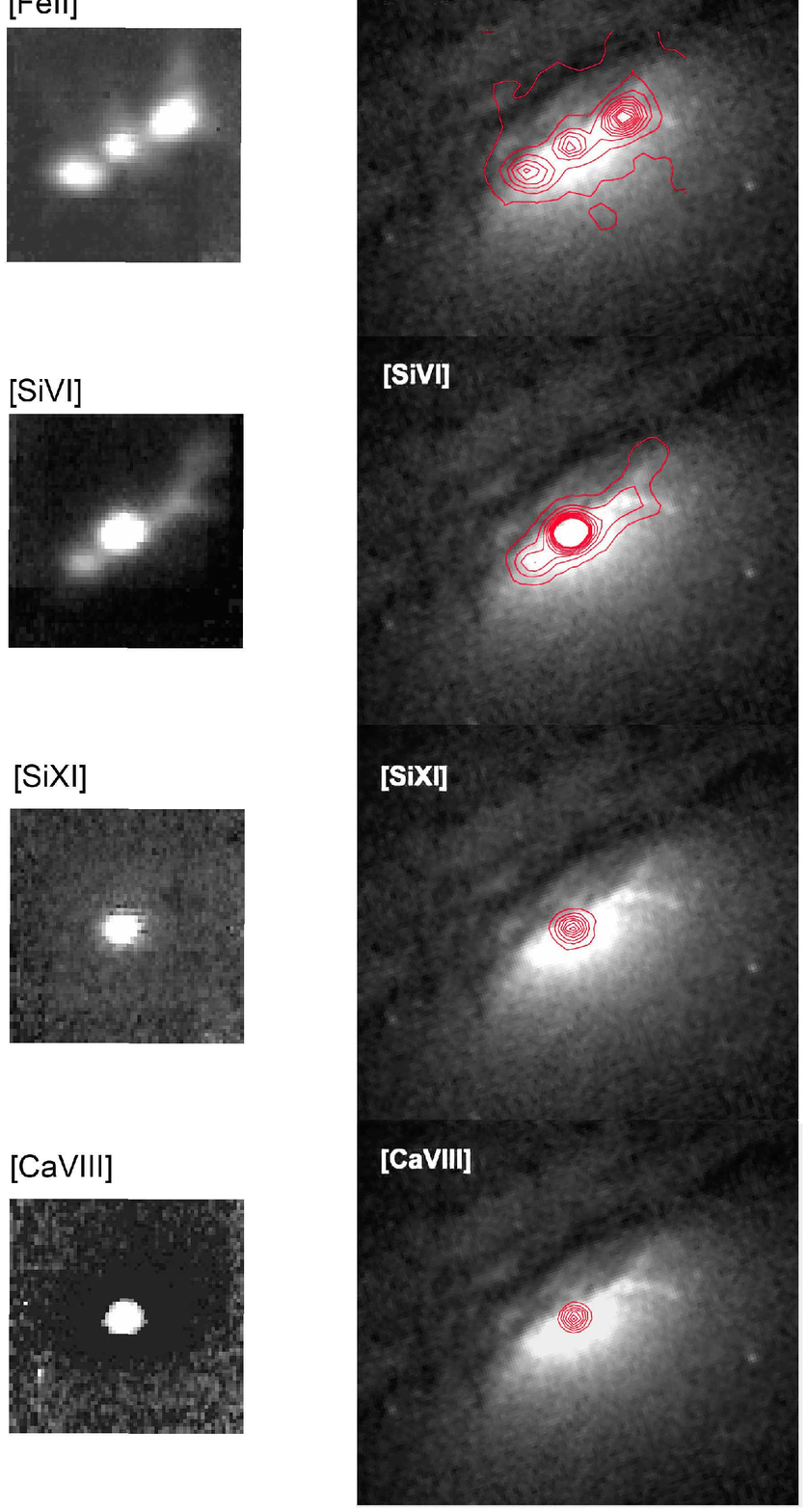} \\
  {FIG.}~\ref{fig:nir_contours} $-$ continued.
\end{figure*}
\begin{figure*}[h!]
\centering
\includegraphics[height=11.5 cm, width=16cm ]{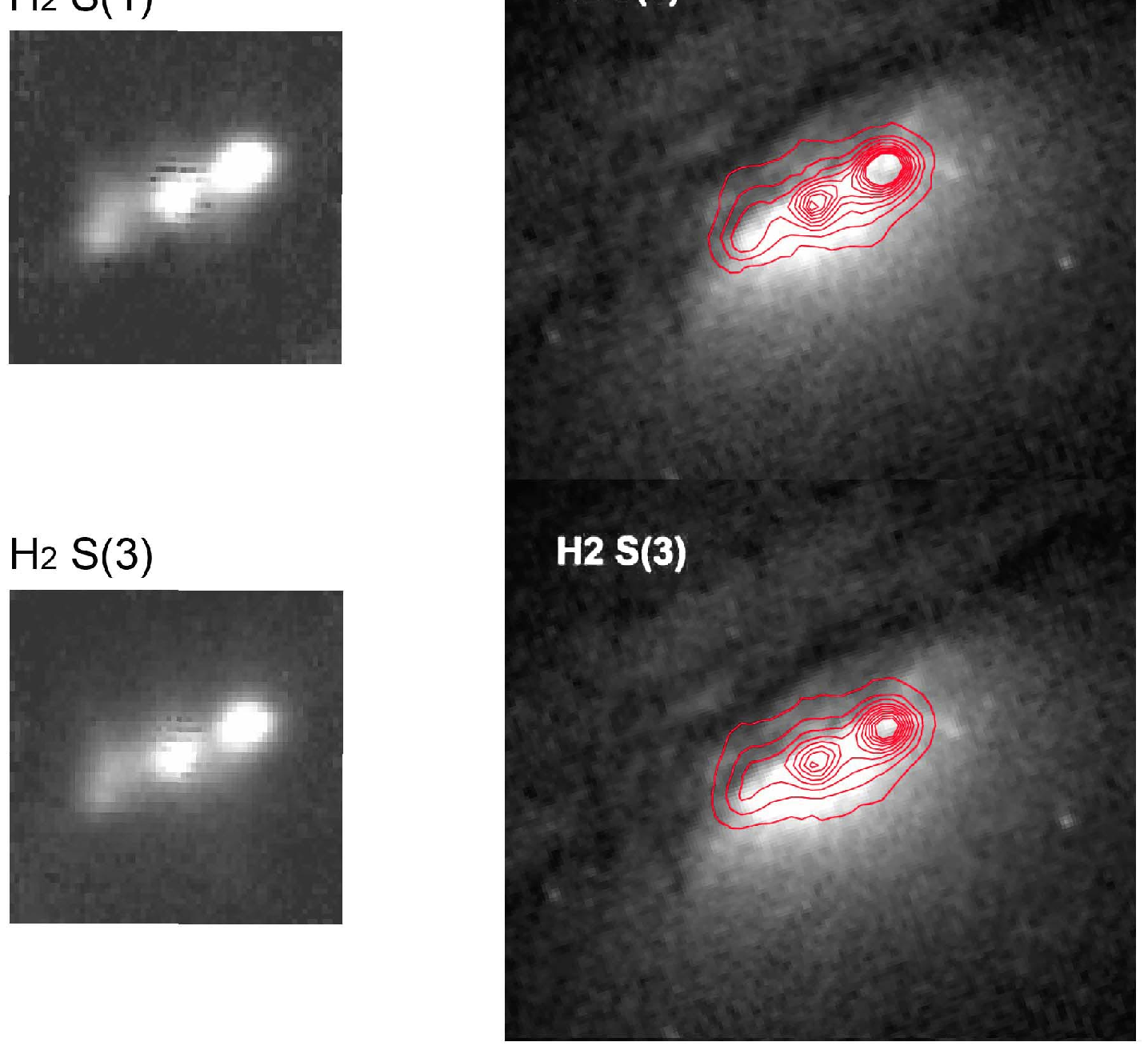} \\
  {FIG.}~\ref{fig:nir_contours} $-$ continued.
\end{figure*}

\begin{figure*}[h!]
\centering
 \includegraphics[width=0.85\textwidth]{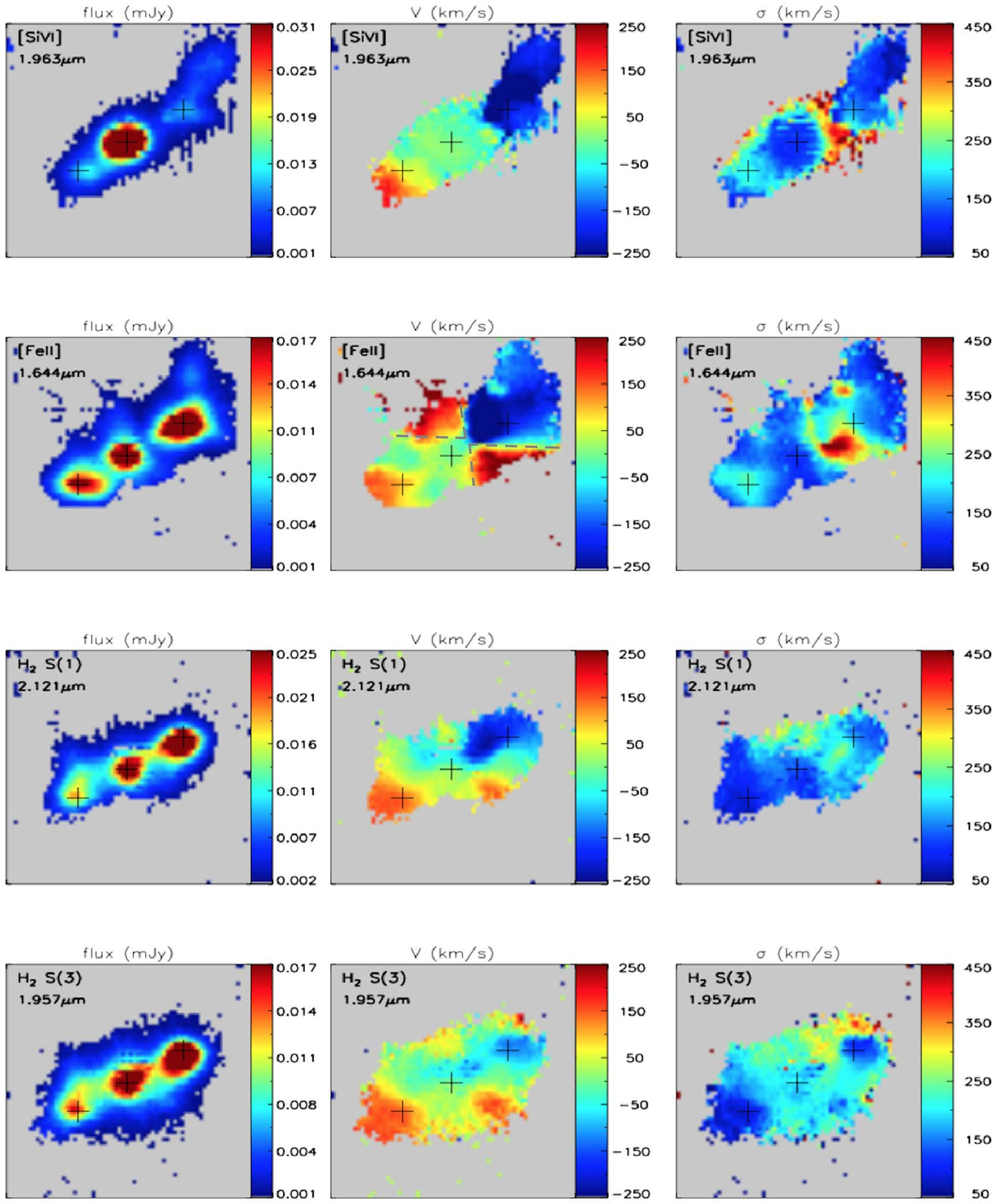} \\
 {FIG.}~\ref{fig:fit_results} $-$ continued.  
\end{figure*}

\begin{figure*}[h!]
 \begin{center}
 \includegraphics[width=0.82\textwidth, height=1.2\textwidth]{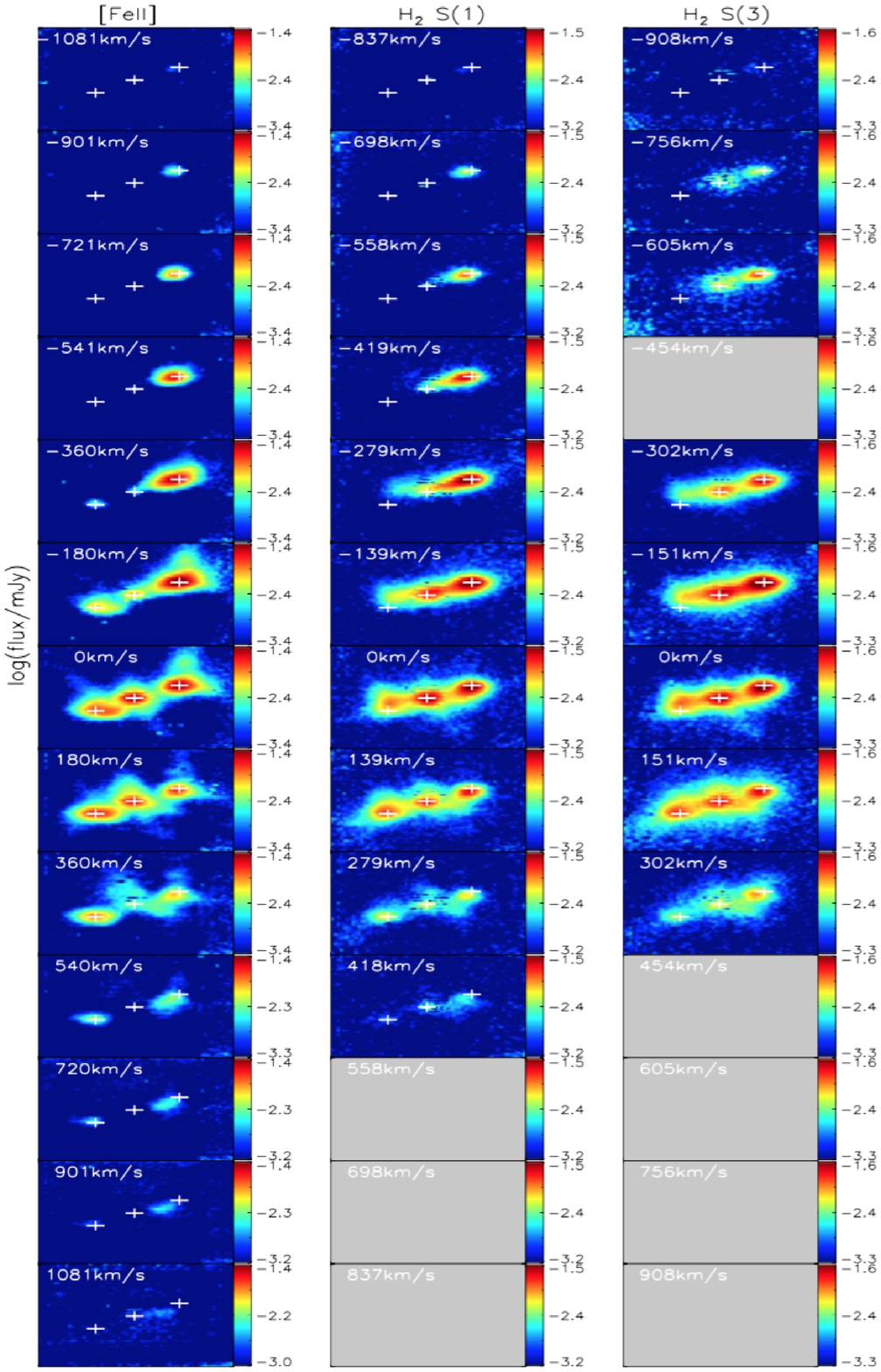} 
\end{center}
{FIG.~\ref{fig:kinematics} $-$ continued.}
\end{figure*}

\clearpage

\end{document}